\documentclass[journal]{IEEEtran}

\ifCLASSINFOpdf
\else
   \usepackage[dvips]{graphicx}
\fi
\usepackage{url}
\usepackage{cite}

\usepackage{siunitx}
\usepackage[colorlinks,urlcolor=blue,linkcolor=blue,citecolor=blue,breaklinks=true]{hyperref}
\usepackage{lipsum}
\usepackage{amsmath}
\usepackage{amsfonts}

\usepackage{bm}
\usepackage{algorithm}
\usepackage{algpseudocode}
\usepackage{graphicx}
\usepackage{tabularray}
\usepackage{threeparttable}
\usepackage{graphicx}
\usepackage{float}
\usepackage{subfig}
\usepackage{xcolor}
\usepackage{environ}
\NewEnviron{revision}{\textcolor{black}{\BODY}}

\begin{document}

\title{Credible Uncertainty Quantification under Noise and System Model Mismatch}

\author{Penggao Yan, Xingqun Zhan, \IEEEmembership{Senior Member, IEEE}, Rui Sun, \IEEEmembership{Member, IEEE}, and Li-Ta Hsu*, \IEEEmembership{Senior Member, IEEE}
\thanks{This work was supported by the National Natural Science Foundation of China (NSFC)/ Research Grants Council (RGC) of Hong Kong Joint Research Scheme under Grant 42561160140 and N\_PolyU502/25. \textit{(Corresponding author: Li-Ta Hsu)}. }
\thanks{Penggao Yan and Li-Ta Hsu are with the Department of Aeronautical and Aviation Engineering, Faculty of Engineering, Hong Kong Polytechnic University, Hong Kong (e-mail:\href{lt.hsu@polyu.edu.hk}{lt.hsu@polyu.edu.hk}); Xingqun Zhan is with the School of Aeronautics and Astronautics, Shanghai Jiao Tong University, Shanghai 200240, China; Rui Sun is with the College of Civil Aviation, Nanjing University of Aeronautics and Astronautics, Nanjing 211106, China.}}

\markboth{Journal of \LaTeX\ Class Files, Vol. xx, No. x, August xxxx}
{Shell \MakeLowercase{\textit{et al.}}: Bare Demo of IEEEtran.cls for IEEE Journals}
\maketitle

\begin{abstract}
State estimators often provide self-assessed uncertainty metrics, such
as covariance matrices, whose credibility is critical for downstream
tasks. However, these self-assessments can be misleading due to
underlying modeling violations like noise model mismatch (NMM) or system model misspecification (SMM). This work addresses this problem by developing a
unified, multi-metric framework that integrates noncredibility index (NCI), negative log-likelihood (NLL), and energy score (ES) metrics, featuring an empirical location test (ELT) to detect system model bias and a directional probing technique that uses the metrics' asymmetric sensitivities to distinguish NMM from SMM. Monte Carlo simulations reveal that the proposed method achieves excellent diagnosis accuracy ($80-100\%$) and significantly outperforms single-metric diagnosis methods. In addition, the parameter sensitivity and the scalability against the state dimension are analyzed using the same simulated dataset. The effectiveness of the proposed method is further validated on a real-world ultra-wideband (UWB) positioning dataset. Finally, the computational complexity of the proposed framework is discussed, providing insights for practical implementation. This framework provides a useful tool for turning patterns
of credibility indicators into actionable diagnoses of model
deficiencies.
\end{abstract}


\IEEEpeerreviewmaketitle

\section{Introduction}

\IEEEPARstart{S}{tate} estimation problems are widely studied in the tracking, navigation, and control community \cite{dubey2023tracking,chen2024kalman,gao2019two,wang2023robust}. Estimators routinely accompany point estimations
with self-assessed uncertainty. For example, Kalman filters (KF) provide covariance matrices, while methods like particle filters produce full predictive distributions. These
self-assessments are informative but rest on modeling assumptions that
may be violated in practice \cite{haggag2022credible,dunik2015random, xia2021fine}. For example, Chauchat et al. \cite{chauchat2021robust} highlighted that the optimality of Kalman filters relies on perfect system modeling, which rarely holds in real-world scenarios. Similarly, Ge et al. \cite{ge2016performance} analyzed the performance degradation of Kalman filters when noise covariances are mismatched. Fortunati et al. \cite{fortunati2017performance} discussed the fundamental performance bounds of parameter estimation under misspecified models, highlighting the impact of modeling violations such as noise-model mismatch (NMM) (e.g., pessimism or optimism) and system-model misspecification (SMM) \cite{gao2019two,li2016maneuvering}. 
This raises a practical question: can the self-assessment be trusted, to what degree, and in which direction of noncredibility (optimism vs.~pessimism)? Following \cite{li2011evaluation}, we refer to
this as the credibility problem. 

In state estimation, credibility is most often judged by a single
statistic, typically the normalized estimation error squared (NEES) and its variants, such as average NEES (ANEES) and noncredibility index
(NCI)\cite{dubey2023tracking,chen2024kalman,li2016maneuvering}. By testing the
agreement between estimation errors and the nominal covariance, NEES
primarily assesses credibility from the calibration perspective, i.e., the consistency between predicted uncertainty and actual outcomes \cite{li2006measuring}. For example, Blasch et al. \cite{blasch2010multitarget} investigated the use of NCI and ANEES for nonlinear estimation performance analysis. They used a nonlinear estimation framework to compare filters like UKF and PF by contrasting credibility metrics against absolute RMS errors. Zhang et al. \cite{zhang2023joint} adopted the NCI to quantify the noncredibility of estimation and decision in joint tracking and classification problems. However, prior work has demonstrated that relying on a single metric can be misleading when model assumptions are violated \cite{chen2022generalized, denti2008glucose, walker2025weaknesses}. Single 
metrics often suffer from directional asymmetry (responding differently to over- 
versus under-estimation) and can be overly sensitive to specific modeling choices. These limitations strongly motivate the need for a more comprehensive, multi-metric 
approach to credibility assessment. 

In parallel, the probabilistic-forecasting and ML/DL communities conduct multi-criteria credibility evaluation \cite{bhatt2022f,chai2018conditional,dang2023differential,chen2022vehicle}, most of which are developed based on the theory of proper scoring rules \cite{gneiting2007strictly,gneiting2014probabilistic}. Notably, the negative log-likelihood (NLL) and energy scores (ES) are widely used to assess calibration and sharpness simultaneously \cite{gneiting2014probabilistic}. For example, Ashok et al. \cite{ashok2023tactis} utilized NLL to validate the TACTiS-2 model, ensuring the learned distribution matches the true distribution. Additionally, Al-Gabalawy et al. \cite{al2021probabilistic} applied NLL minimization to train various deep learning probabilistic models for energy time series, emphasizing the importance of proper scoring rules for calibrated predictions. These practices align with credibility evaluation in state-estimation problems.

To tackle the limitations of single-metric-based diagnosis, we propose a unified multi-metric credibility evaluation framework that integrates NCI, NLL, and ES. Specifically, we first construct an empirical location test (ELT) based on energy distance to statistically detect the presence of SMM. If SMM is detected, we mitigate its impact by centering the estimation errors, thereby isolating the potential NMM effects. Subsequently, we employ a directional probing technique that artificially scales the covariance to generate ``probes'' of NLL and ES. By analyzing the asymmetric responses of these probes, we can robustly distinguish between optimism, pessimism, and SMM. This procedure effectively transforms complex patterns of multiple metrics into actionable diagnoses of model deficiencies.

The proposed method is evaluated on two types of experiments, including Monte Carlo simulations and a real-world ultra-wideband (UWB) positioning dataset. The simulation results cover six distinct credibility scenarios, revealing that the proposed method achieves a diagnosis accuracy of 80\%--100\%, significantly outperforming single-metric baselines. Furthermore, the evaluation on the UWB dataset demonstrates the practical applicability of the framework, where it successfully identifies the coexistence of pessimism and SMM in static positioning periods, a nuance that conventional NEES and NCI methods fail to capture. \begin{revision}This framework provides a practical tool for turning patterns of credibility indicators into actionable diagnoses of model deficiencies, potentially serving as a validation utility for both classical filtering and deep learning-based estimation methods\end{revision}. The contributions of this work are threefold:
\begin{enumerate}
    \item We analytically and experimentally reveal the complementary directional asymmetries of NLL and ES, serving as the theoretical foundation for distinguishing different types of noncredibility.
    \item We propose a unified credibility diagnosis framework featuring an ELT for SMM detection and a directional probing mechanism, which provides a robust solution for disentangling NMM and SMM.
    \item We experimentally demonstrate the effectiveness and superiority of the proposed framework in both controlled simulations and real-world UWB positioning scenarios.
\end{enumerate}

The rest of this article is organized as follows. Section \ref{sec: metrics} analyzes the properties of NEES, NCI, NLL, and ES, pointing out their individual limitations. Section \ref{sec:method} details the proposed unified credibility diagnosis scheme, including the ELT and directional probing techniques. In Section \ref{sec:simulation}, we examine the performance of the proposed method through Monte Carlo simulations. In addition, the parameter sensitivity and the scalability against the state dimension are analyzed. In Section \ref{sec:uwb}, we validate the effectiveness of the framework using the real-world UWB dataset. Section \ref{sec:computation} discusses the computational complexity of the proposed method. Finally, Section \ref{sec:conclusion} gives a summary.

\section{Credibility Metrics Analysis}\label{sec: metrics}  

Let the estimatee and its estimate be $x$ and $\hat{x}$. Define the estimation error $e=x-\hat{x}$ with mean $\mu$ and covariance $\Sigma$. The mean-square error (MSE) is $\mathcal{M}=\mathbb{E}[ee^\top]=\Sigma+\mu\mu^\top$ (equals $\Sigma$ only when $\mu=0$). The estimator reports covariance $\hat{\Sigma}$ and (optionally) MSE $\hat{\mathcal{M}}$. For Monte-Carlo (MC) experiments, the $k$-th run uses $(x_k,\hat{x}_k,e_k,\hat{\Sigma}_k)$ and we run $N$ independent trials. The predictive cumulative distribution function (CDF)/probability density function (PDF) are denoted $\hat{F}(\cdot)$ and $\hat{f}(\cdot)$, respectively.\par
To provide a concrete foundation for the credibility analysis, we consider three representative scenarios in state estimation:
\begin{itemize}
    \item \textbf{Credible}: \((\hat{\Sigma}_k=\Sigma_k,\mu_k=0)\)
    \item \textbf{NMM}: we characterize the mismatch as an incorrect scaling of the covariance, formalized as \(\hat{\Sigma}_k = \rho\,\Sigma_k\) with \(\mu_k = 0\), where \(\rho > 0\) denotes the scaling factor
    \item \textbf{SMM}: we characterize the mismatch as the presence of a constant estimation bias, expressed as \(\mu_k = b_k \neq 0\) with \(\hat{\Sigma}_k = \Sigma_k\), where \(b_k\) is the bias vector.
\end{itemize}
These canonical cases serve as the basis for systematically evaluating the behavior and diagnostic power of various credibility metrics in subsequent sections.

\subsection{Normalized Estimation Error Squared (NEES)}\label{nees}

\begin{equation}
\epsilon_k \;=\; e_k^\top \hat{\Sigma}_k^{-1} e_k.
\label{eq:nees}
\end{equation} 
Consider \(e_k\sim\mathcal N(\mu_k,\Sigma_k)\) ,
we have \begin{equation}
\mathbb{E}[\epsilon_k] \;=\; \operatorname{tr}(\hat{\Sigma}_k^{-1}\Sigma_k)\;+\;\mu_k^\top \hat{\Sigma}_k^{-1}\mu_k.
\end{equation} 
\textbf{Properties.}
\begin{itemize}
    \item \textbf{In the credible case:} \(\mathbb{E}[\epsilon_k]=d\) (i.e., state dimension).
    \item \textbf{In the NMM case with incorrect scaling covariance:} \(\mathbb{E}[\epsilon_k]=d/\rho\).  Define the deviation from the expected value as follows:
    \begin{equation}
    D_k \;=\; \mathbb{E}[\epsilon_k]-d \;=\; d\!\left(\tfrac{1}{\rho}-1\right) \,.
    \end{equation} 
    Notably, $D_k$ is positive for optimism \((\rho<1)\) and
    negative for pessimism \((\rho>1)\). Moreover,
    \(|D_k(\rho)|<|D_k(1/\rho)|\) for \(\rho>1\), i.e., NEES penalizes
    optimism more severely than pessimism.
    \item \textbf{In the SMM case with constant estimation bias:} \(\mathbb{E}[\epsilon_k] = d+ \mu_k^T\Sigma_k^{-1}\mu_k\), and
    \(D_k = \mu_k^T\Sigma^{-1}\mu_k\) is always positive and increases with
    the bias magnitude. Therefore, NEES always penalizes SMM. However, it is difficult to distinguish the SMM and optimism, as in both cases
    \(D_k\) is positive.
\end{itemize}

\subsection{Noncredibility index (NCI)}\label{noncredibility-index-nci}

\begin{equation}
NCI(\{\hat{x}_k\}) = \frac{10}{N}\sum_{k=1}^N \log_{10}(\epsilon_k) - \frac{10}{N}\sum_{k=1}^N \log_{10}(\epsilon_k^*) \,,
\end{equation} 
where \(\epsilon_k^*\) is the NEES of a perfectly credible estimator, calculated as \(\epsilon_k^*= e_k^T \mathcal{M}_k^{-1} e_k\). The magnitude of the NCI directly measures the level of noncredibility.

\textbf{Properties.}
\begin{itemize}
    \item \textbf{In the credible case:} NCI is zero.
    \item \textbf{In the NMM case with incorrect scaling covariance:} The NCI is given by
    \begin{equation}
    NCI(\{\hat{x}_k\},\rho)= - \frac{10}{N}\sum_{k=1}^N \log_{10} \rho.
    \end{equation} 
    This value is positive for optimism (\(\rho<1\)) and negative for pessimism (\(\rho>1\)). As \(|NCI(\{\hat{x}_k\},\rho)|=|NCI(\{\hat{x}_k\},\frac{1}{\rho})|\), NCI penalizes optimism and pessimism equally.
    \item \textbf{In the SMM case with constant estimation bias:} The NCI is always non-negative (see Appendix \ref{appendix-nci}). Therefore, NCI always penalizes SMM. However, similar to NEES, it cannot distinguish SMM from optimism, as both result in non-negative NCI values.
\end{itemize}

\subsection{Negative Log-Likelihood
(NLL)}\label{negative-log-likelihood-nll}

\begin{equation}
\mathrm{NLL}(\hat{F}_k,x_k) = -\log \hat{f}_k(x_k) \,.
\label{eq:nll}
\end{equation} 
Since \(\hat{f}_k(x_k)\le 1\), NLL is non-negative. Consider
\(\hat{F}_k=\mathcal N(\hat{x}_k,\hat{\Sigma}_k)\) and
\(e_k=x_k-\hat{x}_k\sim\mathcal N(\mu_k,\Sigma_k)\) , we have 
\begin{small}
\begin{equation}
\mathbb{E}[\mathrm{NLL}] \;=\; \tfrac12\!\left(\operatorname{tr}(\hat{\Sigma}_k^{-1}\Sigma_k) + \mu_k^\top \hat{\Sigma}_k^{-1}\mu_k + \ln|\hat{\Sigma}_k| + d\ln(2\pi)\right).
\end{equation} 
\end{small}

\textbf{Properties}. 
\begin{itemize}
\item \textbf{In the NMM case with incorrect scaling covariance:} 
\begin{equation}
  \mathbb{E}[\mathrm{NLL}] = \tfrac12\!\left(\tfrac{d}{\rho}+ d\ln\rho + \ln|\Sigma_k| + d\ln(2\pi)\right) \,.
  \label{eq:exp_nll_1} 
\end{equation}
Appendix \ref{appendix-nll} shows that
$|\mathbb{E}[\mathrm{NLL}](\rho)| < |\mathbb{E}[\mathrm{NLL}](1/\rho)|$, indicating that the NLL penalizes optimism more severely than pessimism. This pronounced sensitivity to optimism, which we will later exploit, makes it a powerful tool for diagnosing model misspecification.

\item \textbf{In the SMM case with constant estimation bias:}
\begin{equation}
  \mathbb{E}[\mathrm{NLL}] \;=\; \tfrac12\!\left(d+ \mu_k^\top \Sigma^{-1}_k\mu_k + \ln|\Sigma_k| + d\ln(2\pi)\right),
  \end{equation}
  revealing that the NLL increases with the bias magnitude. Therefore, the NLL always
  penalizes SMM. Since the NLL always takes a non-negative value, it is difficult to distinguish NMM and SMM solely by using the NLL.
\end{itemize}

\subsection{Energy Score (ES)}\label{energy-score-es}

\begin{equation}
\mathrm{ES}(\hat{F}_k,x_k)=\mathbb{E}_{Y\sim \hat{F}_k}\|Y-x_k\|_2-\tfrac12\,\mathbb{E}_{Y,Y'\sim \hat{F}_k}\|Y-Y'\|_2.
\label{eq:es}
\end{equation} 
The first term measures calibration (distance to truth), and the second term measures sharpness (concentration). The ES is always non-negative (see Appendix \ref{appendix-es}). To compute ES, we approximate the expectation operation using Monte Carlo sampling with M samples ($M=500$ is used in this work).

\begin{figure}[!htb]
  \centering
\subfloat[~~~~]{%
     \includegraphics[width=75mm]{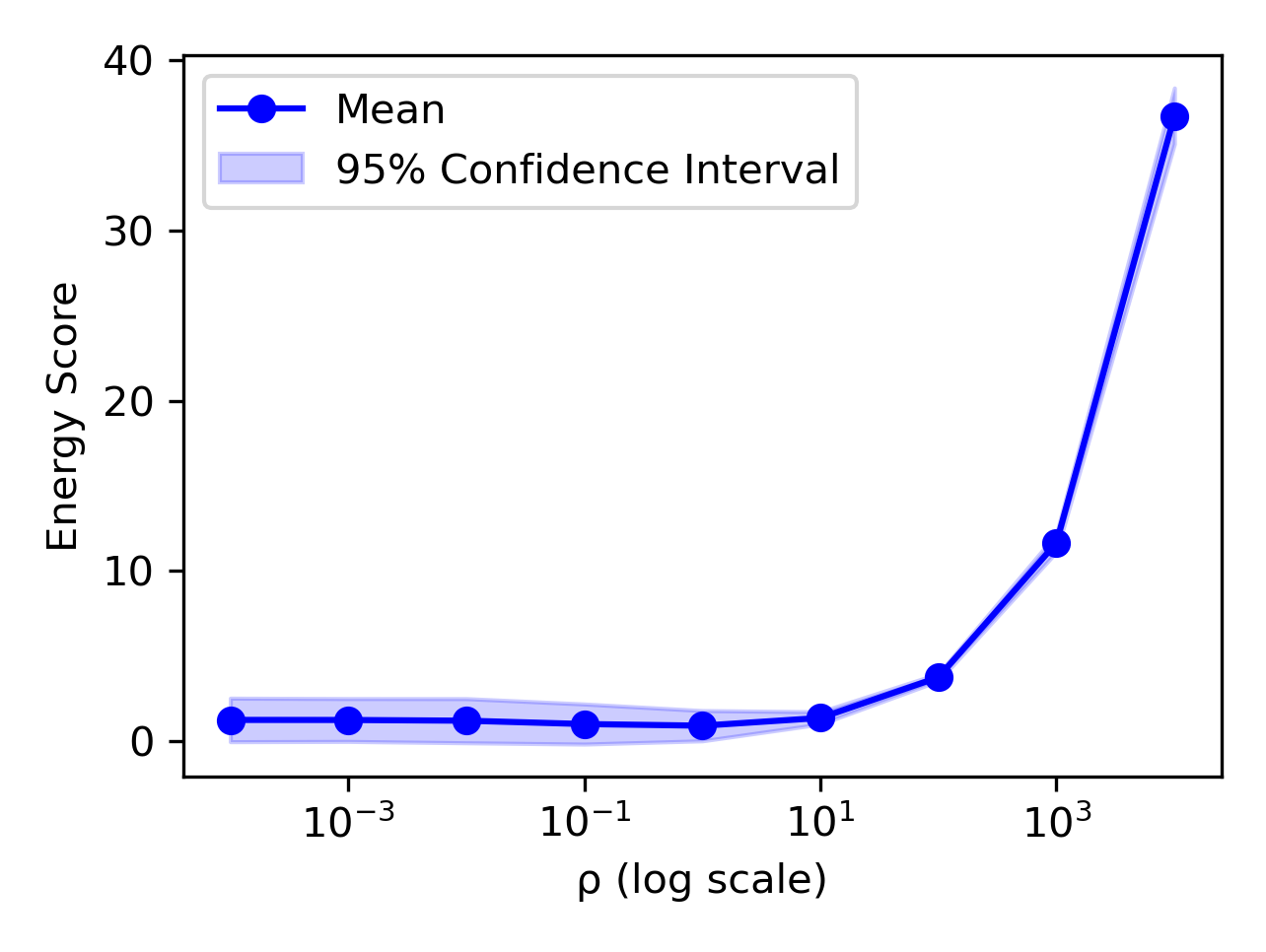}}
  \\
\subfloat[~~~~]{%
      \includegraphics[width=75mm]{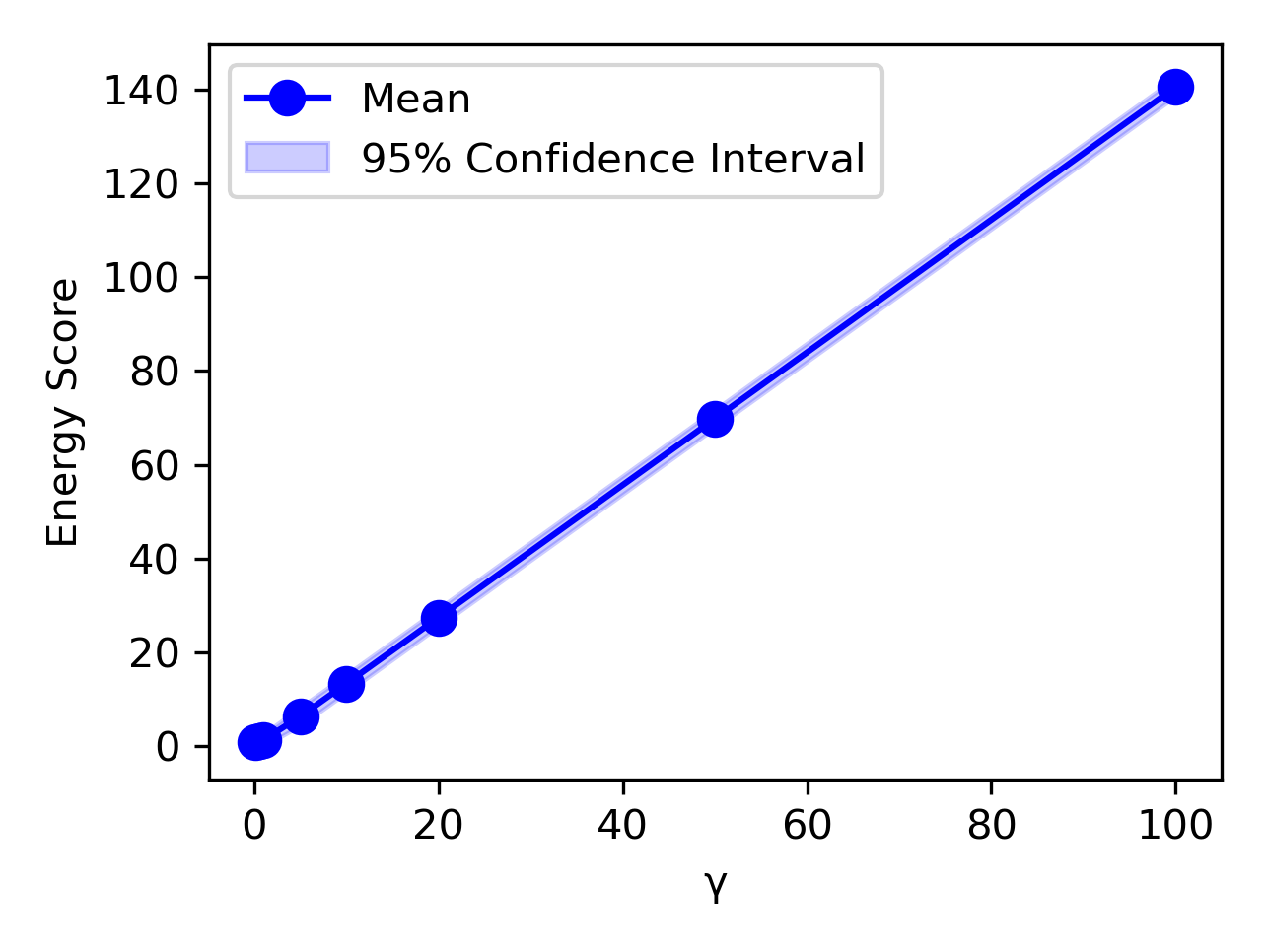}}
\caption{Monte-Carlo simulations show how ES varies against (1) $\rho$ and (2) $\gamma$.}
\label{fig:ES_scale_bias}
\end{figure}

\textbf{Properties}. Due to the complexity of the ES formulation, there are
no analytical forms of ES and its expectation. Therefore, we implement
Monte-Carlo simulations to study the sensitivity of ES. Specifically, we
study the case of a multivariate normal distribution
\(\hat{F}_k = \mathcal{N}(\hat{x}_k,\hat{\Sigma}_k)\) with
\(\hat{x}_k\sim \mathcal{N}(x_k+\mu_k,\Sigma_k)\).

\begin{itemize}
\item
\textbf{In the NMM case with incorrect scaling covariance:}  We set \(x_k=\mathbf{0}\) and \(\Sigma_k = \rm{I}_2\). We
  calculate the sample mean of ES by implementing 5,000
  Monte Carlo runs for each \(\rho\). Fig. \ref{fig:ES_scale_bias}a plots the sample mean of
  ES against \(\rho\). It is evident that the ES penalizes pessimism
  more severely than optimism, a characteristic that stands in direct
  contrast to the properties of the NLL. This contrasting sensitivity to pessimism is fundamental to the proposed unified diagnostic approach, as it provides complementary information to the NLL. 
\item
\textbf{In the SMM case with constant estimation bias:} We set
  \(\mu_k=\gamma[1,1]^T\), \(x_k=\mathbf{0}\), \(\Sigma_k = \rm{I}_2\).
  Similarly, we implement the Monte Carlo simulation (5,000 runs for
  each \(\gamma\)) and plot the relationship between the sample mean of ES
  and \(\gamma\) in Fig. \ref{fig:ES_scale_bias}b. Evidently, the ES always penalizes SMM. Since the ES always takes a non-negative value, it
  is difficult to distinguish NMM and SMM solely by using the ES.
\end{itemize}

Table \ref{tab:metric_summary} summarizes the properties of NEES, NCI, NLL, and ES. While each of the metrics discussed provides a unique perspective on credibility, each metric has its own limitations. NEES and NCI struggle to distinguish optimism from SMM, while NLL and ES show opposite sensitivities to covariance scaling but cannot independently identify the direction of noncredibility. This motivates our development of a unified evaluation scheme that synergistically combines these metrics to provide a more complete and reliable diagnosis.

\begin{table*}[!htb]
  \centering
  \caption{Summary of Credibility Metrics and Their Properties}
  \label{tab:metric_summary}
  \begin{footnotesize}
  \begin{tblr}{
    row{1} = {font=\bfseries},
    hline{1,6} = {-}{1.5pt},
    hline{2} = {-}{0.75pt},
    colspec = {Q[40pt]Q[50pt]Q[150pt]Q[170pt]},
    width = \textwidth,
  }
  Metric & Credible Case & NMM Case (Scaling $\rho$) & SMM Case (Bias $\mu_k$) \\
  \textbf{NEES}   & $\mathbb{E}[\epsilon_k] = d$ & $\mathbb{E}[\epsilon_k] = d/\rho$. Penalizes optimism ($\mathbb{E}[\epsilon_k]>d$) more severely than pessimism ($\mathbb{E}[\epsilon_k]<d$). &$\mathbb{E}[\epsilon_k]>d$. $\mathbb{E}[\epsilon_k]$ increases with the bias magnitude. Difficult to distinguish SMM and optimism. \\
  \textbf{NCI}    & $NCI = 0$ & Symmetric penalty for optimism ($NCI>0$) and pessimism ($NCI<0$). & $NCI \geq 0$. Difficult to distinguish SMM and optimism. \\
  \textbf{NLL}    & Minimized & Penalizes optimism more severely than pessimism. In both cases, $NLL>0$. & $NLL>0$. Increases with bias magnitude. \\
  \textbf{ES}     & Minimized & Penalizes pessimism more severely than optimism. In both cases, $ES>0$. & $ES>0$. Increases with bias magnitude. \\
  \end{tblr}
\end{footnotesize}
\end{table*}

\section{Unified Credibility Diagnosis Scheme}\label{sec:method}

\begin{figure}[!htb]
    \centering
    \includegraphics[width=\linewidth]{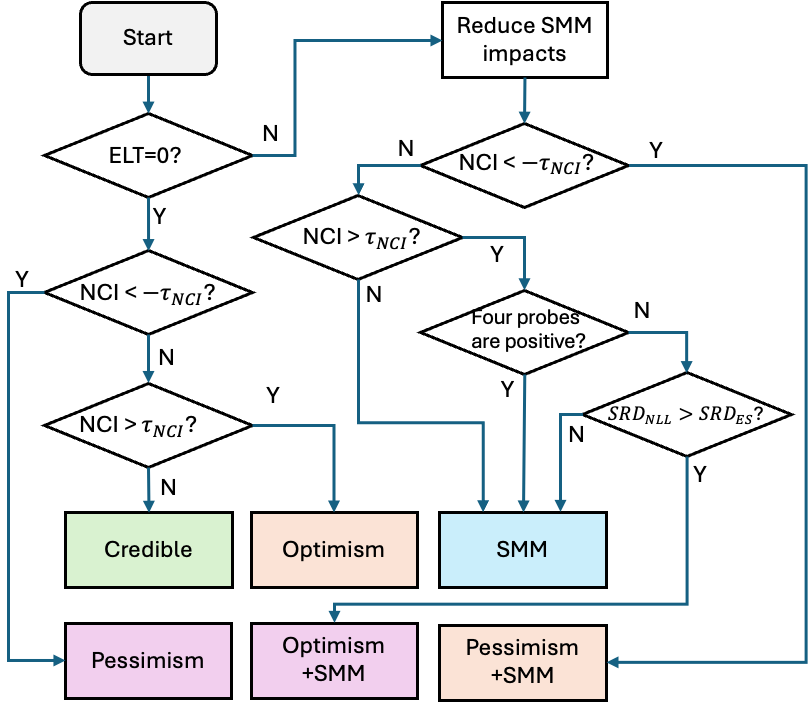}
    \caption{The flowchart of the proposed unified credibility diagnosis method.}
    \label{fig:flowchart}
\end{figure}

We propose a heuristic procedure to distinguish NMM (optimism or pessimism) and SMM. A heuristic framework is chosen because a purely analytical solution is not straightforward, and the combined effects of SMM and NMM are difficult to formally disentangle. Our step-by-step procedure is therefore designed to navigate this complexity. Fig. \ref{fig:flowchart} gives the flowchart of the proposed algorithm. 

\subsection{Identify the impacts of SMM}
As discussed in Section \ref{sec: metrics}, standard metrics such as NEES, NCI, NLL, and
ES are unable to distinguish between NMM and SMM due to their inherent formulations. To overcome
this limitation, we construct an ELT test based on energy distance \cite{szekely2013energy} to first identify and remove the impacts of SMM, disentangling the combined effects of SMM and NMM. 

The energy distance quantifies the dissimilarity between two probability distributions \(F\) and \(G\), and is defined as: 
\begin{equation}
  \begin{aligned}
\mathcal{D}_\alpha^2(F, G) & = 2\,\mathbb{E}_{X \sim F,\, Y \sim G} \|X - Y\|^\alpha \\
  &- \mathbb{E}_{X, X' \sim F} \|X - X'\|^\alpha \\ 
  &- \mathbb{E}_{Y, Y' \sim G} \|Y - Y'\|^\alpha,
\end{aligned}
\end{equation} 
where \(\alpha \in (0, 2]\). In this work, we set \(\alpha = 1\), \begin{revision}which corresponds to the standard Energy Distance \cite{szekely2013energy}.\end{revision}

To test for SMM, we examine the distribution
of the whitened estimation errors, \(s_k = \hat{\Sigma}_k^{-1/2} e_k\).
In the absence of SMM, the distribution of
\(\{s_k\}\) should be symmetric about the origin, i.e., identical to its
mirror image \(\{-s_k\}\). This leads to the following hypothesis test:
\begin{equation}
  \begin{aligned}
& H_0: \mathcal{L}(\{s_k\}) = \mathcal{L}(\{-s_k\}) \\
& H_1: \mathcal{L}(\{s_k\}) \neq \mathcal{L}(\{-s_k\}),
\end{aligned}
\end{equation} 
where \(\mathcal{L}(\cdot)\) denotes the empirical distribution. An ELT test is then constructed. Specifically, a test statistic is constructed as the energy distance between the empirical distributions of \(\{s_k\}\) and \(\{-s_k\}\):
\begin{equation}
\begin{aligned}
T_{\text{obs}} &= \mathcal{D}_\alpha^2\big(\mathcal{L}(\{s_k\}),\, \mathcal{L}(\{-s_k\})\big) 
\\
&= \frac{2}{N(N-1)} \sum_{1 \leq i < j \leq N} \left( \|s_i + s_j\|^\alpha - \|s_i - s_j\|^\alpha \right),
\end{aligned}
\label{eq:energy_distance}
\end{equation} 
where \(N\) is the number of samples. When the distribution of
estimation errors is centered at the origin, the average pairwise sum
and difference distances are balanced, resulting in a small value of
\(T_{\text{obs}}\). However, if the distribution is systematically
shifted away from the origin (indicative of SMM), the
average \(\|s_i + s_j\|\) increases, leading to a larger value of the
test statistic.

To formally assess the statistical significance of the observed energy distance, we employ a sign-flip randomization test. This test is based on the null hypothesis \(H_0\) of central symmetry. Specifically, let \(\xi_k \in \{-1, +1\}\) be independent Rademacher random variables. Under \(H_0\), the distribution of the sign-flipped set \(\{\xi_1 s_1, \dots, \xi_N s_N\}\) is identical to that of the original set \(\{s_1, \dots, s_N\}\):
\begin{equation}
\{\xi_1 s_1, \dots, \xi_N s_N\} \stackrel{d}{=} \{s_1, \dots, s_N\}.
\end{equation}

Therefore, by applying random sign flips to the whitened errors \(\{s_k\}\), we can generate samples from the null distribution of the test statistic. The conditional distribution of the energy distance statistic \(\mathcal{D}_\alpha^2\big(\mathcal{L}(\{\xi_k s_k\}), \mathcal{L}(\{-\xi_k s_k\})\big)\) serves as the exact reference distribution for our hypothesis test, allowing us to calculate a p-value.

In practice, we perform the following randomization procedure to
construct this null distribution: for each iteration \(b = 1, \dots, B\)
(usually \(B\geq 1000\)), we independently sample Rademacher random
variables \(\xi_k^{(b)} \in \{\pm 1\}\) for each \(k\), apply the sign
flips to obtain \(\{\xi_k^{(b)} s_k\}\), and then compute the
corresponding randomized test statistic by the energy distance: \begin{equation}
T^{(b)} = \mathcal{D}^2_\alpha(\{\xi_k^{(b)} s_k\},\{-\xi_k^{(b)} s_k\}).
\end{equation}

The (one-sided) randomized \(p\)-value is then estimated as \begin{equation}
p_{\text{ELT}} = \frac{1 + \#\{b : T^{(b)} \geq T_{\text{obs}}\}}{B + 1}.
\end{equation}

Finally, we define the ELT decision as \begin{equation}
\mathrm{ELT} = \mathbf{1}\{p_{\text{ELT}} < \alpha_{\text{sig}}\},
\end{equation} where \(\alpha_{\text{sig}}\) is the significance level. An outcome
of \(\mathrm{ELT} = 1\) indicates statistical evidence for SMM.

\subsection{Evaluation of noncredibility direction without SMM}

When \(ELT=0\), we declare that SMM does not
exist. The remaining question is whether the estimation is pessimistic,
optimistic, or credible. To answer this question, we simply use the NCI
metric due to its ability to evaluate the noncredibility direction.
Specifically, we define a positive threshold \(\tau_{\text{NCI}}>0\). If
NCI is smaller than \(-\tau_{\text{NCI}}\), the estimation is said to be
pessimistic; If NCI is larger than \(\tau_{\text{NCI}}\), the estimation
is said to be optimistic; Otherwise, the estimation is said to be
credible.

\subsection{Evaluation of noncredibility direction with SMM}\label{evaluation-of-noncredibility-direction-with-system-model-misspecification}

\subsubsection{Reduce the impacts of SMM}
When \(ELT=1\), we declare the SMM exists. If
the SMM and NMM both exist,
the effects of SMM will disrupt the assessment
of the direction of noncredibility. One intuitive solution is to
subtract the sample mean of estimation errors from the estimation,
i.e., $\check{x}_k = \hat{x}_k - \frac{1}{N}\sum_{k=1}^N e_k$. The corresponding predictive distribution is given by $
\check{F}_k =\mathcal{N}(\check{x}_k,\hat{\Sigma}_k)$. The modified estimation \(\check{x}_k\) and predictive distribution
\(\check{F}_k\) will be used to further determine whether the estimation
is pessimistic or optimistic.

\subsubsection{Use NCI to tentatively evaluate pessimism}
After mitigating the impact of SMM, we use the NCI to tentatively identify the direction of noncredibility. If \(|\text{NCI}(\{\check{x}_k\})| \leq \tau_{\text{NCI}}\), the adjusted estimation \(\check{x}_k\) can be considered credible in terms of its noise model. We can therefore conclude that the original estimation is only affected by SMM; If \(\text{NCI}(\{\check{x}_k\}) < -\tau_{\text{NCI}}\), this provides strong evidence that the original estimation is affected by both SMM and pessimism; If \(\text{NCI}(\{\check{x}_k\}) > \tau_{\text{NCI}}\), the interpretation is more subtle. A positive NCI may result from either residual SMM or genuine optimism. Therefore, the analysis in Section \ref{sec:probes} is required to distinguish this ambiguity.

\subsubsection{Directional probes using NLL and ES}\label{sec:probes}
In this step, we exploit the asymmetric sensitivities of the NLL and ES to pessimism and optimism. The core idea is to ``probe" the credibility of the estimate by artificially scaling its covariance. By observing how NLL and ES react differently to these optimistic and pessimistic probes, we can infer the underlying nature of the uncertainty. 

To capture the asymmetric sensitivity of NLL and ES to covariance scaling, we construct the following probes of NLL and ES as follows:
\begin{equation}
\begin{aligned}
\Delta^-_{\text{NLL}} &= \mathrm{NLL}(\check{F}_k(1/c), x_k) - \mathrm{NLL}(\check{F}_k, x_k) \\
\Delta^+_{\text{NLL}} &= \mathrm{NLL}(\check{F}_k(c), x_k) - \mathrm{NLL}(\check{F}_k, x_k) \\
\Delta^-_{\text{ES}} &= \mathrm{ES}(\check{F}_k(1/c), x_k) - \mathrm{ES}(\check{F}_k, x_k) \\
\Delta^+_{\text{ES}} &= \mathrm{ES}(\check{F}_k(c), x_k) - \mathrm{ES}(\check{F}_k, x_k)\,,
\end{aligned}
\end{equation}
where $\check{F}_k(c) =\mathcal{N}(\check{x}_k,c\hat{\Sigma}_k)$ is the scaled predictive distribution by the scaling factor \(c>1\).  Importantly, when
\(\check{x}_k\) is free from NMM, all four
probes are expected to be positive, reflecting the fact that both NLL and ES increase
under either pessimistic or optimistic estimation scenarios.

To quantify the asymmetry of NLL and ES's response to optimistic versus pessimistic scaling, we construct the slope relative difference (SRD) as follows:
\begin{equation}
\text{SRD}_{\text{NLL}} = \frac{c|\Delta^-_{\text{NLL}}| - |\Delta^+_{\text{NLL}}|}{|\Delta^+_{\text{NLL}}|}, \text{SRD}_{\text{ES}} = \frac{c|\Delta^-_{\text{ES}}| - |\Delta^+_{\text{ES}}|}{|\Delta^+_{\text{ES}}|} \,,
\end{equation}
which measures the relative difference between an optimistic probe (\(|\Delta^-|\)) and a pessimistic probe (\(|\Delta^+|\)). The factor \(c\) accounts for the unequal step sizes of the probes (\(1 \to 1/c\) vs. \(1 \to c\)), effectively comparing the local slopes of the scoring rule in each direction. Since NLL exhibits stronger sensitivity to optimism
than pessimism, it is expected that
\(\text{SRD}_{\text{NLL}} > \text{SRD}_{\text{ES}}\) when the estimation
is optimism. In contrast, 
\(\text{SRD}_{\text{ES}} > \text{SRD}_{\text{NLL}}\) is expected when
the estimation is pessimism.

Based on the above findings, we propose the following two-step diagnosis procedure. First, we examine the signs of the probes. If all four probes are positive, it suggests that the bias-corrected estimate \(\check{x}_k\) is likely free from NMM, and thus we conclude that the original estimate \(\hat{x}_k\) is only affected by SMM. If any of these four metrics is not positive, it indicates that \(\check{x}_k\) may still exhibit optimism. To resolve this, we proceed to the second step: comparing the SRD values. If \(\text{SRD}_{\text{NLL}} > \text{SRD}_{\text{ES}}\), we confirm \(\check{x}_k\) is optimism, and therefore declare that \(\hat{x}_k\) is affected by both SMM and optimism. Otherwise, the evidence for optimism is not conclusive, and we revert to concluding that \(\hat{x}_k\) is primarily affected by SMM. \par

\subsection{Pseudocode of the proposed Algorithm}\label{app:algorithm}

The pseudocode of the proposed algorithm is listed in Algorithm \ref{alg:1}.

\begin{algorithm}
\caption{Unified Credibility Evaluation Scheme}
\label{alg:1}
\begin{footnotesize}
\begin{algorithmic}[1]
\If{ELT $= 0$}
    \Comment{No SMM}
    \If{NCI $< -\tau_{\text{NCI}}$}
        \State \textbf{output:} pessimism
    \ElsIf{NCI $> \tau_{\text{NCI}}$}
        \State \textbf{output:} optimism
    \Else
        \State \textbf{output:} credible
    \EndIf
\Else
    \Comment{SMM exists}
    \State Subtract mean error from estimation
    \If{NCI $< -\tau_{\text{NCI}}$}
        \State \textbf{output:} pessimism + SMM
    \ElsIf{NCI $> \tau_{\text{NCI}}$}
        \Comment{Maybe also optimism or small pessimism}
        \If{$\Delta^-_{\text{NLL}} > 0$ \textbf{and} $\Delta^+_{\text{NLL}} > 0$ \textbf{and} $\Delta^-_{\text{ES}} > 0$ \textbf{and} $\Delta^+_{\text{ES}} > 0$}
            \Comment{Both metrics increase when moving away}
            \State \textbf{output:} SMM
        \ElsIf{$\text{SRD}_{\text{NLL}} > \text{SRD}_{\text{ES}}$}
            \State \textbf{output:} optimism + SMM
        \Else
            \State \textbf{output:} SMM
        \EndIf
    \Else
        \State \textbf{output:} SMM
    \EndIf
\EndIf
\end{algorithmic}
\end{footnotesize}
\end{algorithm} 

\section{Simulation Experiments}\label{sec:simulation}

\subsection{Experimental Design}\label{a.-experimental-design}

We validate the proposed unified evaluation scheme through Monte Carlo
simulations with controlled synthetic data covering six distinct
credibility scenarios.

\textbf{Setup.} We consider a 2-D state estimation problem
(\(d=2\)) across six scenarios with 50 trials each. Each trial comprises
100 Monte Carlo runs. True states \(x_k\) are generated from a Gaussian distribution with mean \(\mathbf{0}\) and covariance \(\Sigma_{\text{true},k}\), where \(\Sigma_{\text{true},k}\) is randomized per trial via QR decomposition
with eigenvalues uniformly distributed in \([0.5, 2.0]\). State estimates are generated as
\(\hat{x}_k \sim \mathcal{N}(x_k + \mu_k, \Sigma_{\text{true},k})\), and
claimed covariances are
\(\hat{\Sigma}_k = \rho \cdot \Sigma_{\text{true},k}\).

\textbf{Scenarios}. Six scenarios are considered. Except for the credible
scenario, each scenario uses a wide parameter range to examine the algorithm's robustness:

\begin{enumerate}
\item
  \textbf{Credible:} \(\mu_k = \mathbf{0}\), \(\rho = 1\)
\item
  \textbf{Optimism:} \(\mu_k = \mathbf{0}\),
  \(\rho \sim \text{Uniform}[0.1, 0.8]\)
\item
  \textbf{Pessimism:} \(\mu_k = \mathbf{0}\),
  \(\rho \sim \text{Uniform}[1.25, 10]\)
\item
  \textbf{SMM:} \(\mu_k\) with randomized direction and \\
  \(\|\mu_k\| \sim \text{Uniform}[1.6, 2.4]\), \(\rho = 1\)
\item
  \textbf{Optimism + SMM:} \(\mu_k\) with randomized direction and
  \(\|\mu_k\| \sim \text{Uniform}[1.6, 2.4]\),
  \(\rho \sim \text{Uniform}[0.1, 0.8]\)
\item
  \textbf{Pessimism + SMM:} \(\mu_k\) with randomized direction and
  \(\|\mu_k\| \sim \text{Uniform}[1.6, 2.4]\),
  \(\rho \sim \text{Uniform}[1.25, 10]\)
\end{enumerate}

\textbf{Algorithm Parameters.} \(\tau_{\text{NCI}} = 0.5\) dB,
\(\alpha_{\text{sig}} = 0.05\), and directional probe scaling \(c = 2\).

\subsection{Experimental Results}\label{b.-results-and-analysis}
Table \ref{tab:accuracy} compares the classification accuracy of the proposed algorithm with baseline algorithms, including NCI and NEES.  Since NLL and ES cannot independently assess the direction of noncredibility, they are not considered in benchmarking. \begin{revision}Single-metric methods exhibit severe limitations: Although the NEES-based method achieves a remarkable 100.0\% accuracy in ``Optimism", it fails in detecting ``Credible" scenarios (6.0\%). Similarly, the NCI-based method achieves 100.0\% accuracy in ``Pessimism" but only shows 56.0\% accuracy in the ``Optimism" scenarios. Moreover, both single-metric methods fail to detect SMM-related scenarios. This validates the necessity of the multi-metric
approach for comprehensive credibility assessment. The proposed method yields the most balanced results, consistently achieving high accuracy (80.0\%-100.0\%) across all scenarios. \end{revision}

\begin{table}[!htb]
  \centering
  \caption{Diagnosis accuracy of the proposed method and single-metric baselines}
  \label{tab:accuracy}
  \begin{footnotesize}
  \begin{tblr}{
    hline{1,8} = {-}{1.5pt},
    hline{2} = {-}{0.75pt},
  }
  & \textbf{Proposed} & NEES & NCI \\
  Credible        & \textbf{94.0\%} & 6.0\%  & 0\% \\
  Optimism        & 84.0\% & \textbf{100.0\%}  & 56.0\%  \\
  Pessimism       & 90.0\% & 74.0\% & \textbf{100.0\%}  \\
  SMM             & \textbf{80.0\%} & 0.0\%  & 0.0\%  \\
  Optimism+SMM    & \textbf{82.0\%} & 0.0\%  & 0.0\%  \\
  Pessimism+SMM   & \textbf{100.0\%} & 0.0\%  & 0.0\%  
  \end{tblr}
  \end{footnotesize}
\end{table}

Table \ref{tab:confusion_matrix} summarizes the confusion patterns of the diagnosis results
of the proposed method. For ``Optimism+SMM" scenarios, 9 cases are classified
as SMM--not incorrect since ``Optimism+SMM" inherently contains SMM,
but incomplete as it misses the optimism component. Similarly, 10 pure
SMM cases are classified as ``Pessimism + SMM,'' indicating the
algorithm correctly identifies the bias but erroneously detects pessimism in
the bias-corrected estimates. In ``Optimism'' scenarios, 6 cases are
misclassified as ``Credible,'' suggesting the algorithm conservatively requires stronger evidence for optimism detection, while 2 cases trigger
``Optimism + SMM,'' indicating spurious bias detection in purely
optimistic conditions. The confusion patterns demonstrate that
misclassifications often represent partial but meaningful detections
rather than complete algorithmic failures, supporting the framework's
utility for practical credibility assessment.

\begin{table}[!htb]
  \centering
  \caption{Confusion matrix of the diagnosis results of the proposed method}
  \label{tab:confusion_matrix}
  \begin{threeparttable}
  \begin{footnotesize}
  \begin{tblr}{
    cell{1}{1} = {r=2}{},
    cell{1}{2} = {c=6}{c},
    vline{2} = {1-8}{},
    hline{2} = {2-7}{0.75pt},
    hline{3} = {-}{0.75pt},
    hline{1,9} = {-}{1.5pt},
  }
  True Scenario  & Diagnosis Results &          &           &      &       &                \\
                 & Credible  & O & P & SMM & O+SMM & P+SMM \\
  Credible       & 47        & 0        & 0         & 3    & 0             & 0              \\
  O\tnote{1}      & 6         & 42       & 0         & 0    & 2             & 0              \\
  P\tnote{1}      & 1         & 0        & 45        & 0    & 0             & 4              \\
  SMM           & 0         & 0        & 0         & 40   & 0             & 10             \\
  O+SMM  & 0         & 0        & 0         & 9    & 41            & 0              \\
  P+SMM & 0         & 0        & 0         & 0    & 0             & 50             
  \end{tblr}
  \begin{tablenotes}
    \item[1] O: Optimism; P: Pessimism.
  \end{tablenotes}
  \end{footnotesize}
  \end{threeparttable}
\end{table}

\begin{revision}
\subsection{Sensitivity Analysis}\label{sec:sens}
To further validate the performance of the proposed framework, we conduct a sensitivity analysis on its key parameters: the NCI threshold \(\tau_{\text{NCI}}\), the significance level \(\alpha_{\text{sig}}\) for ELT, and the directional probing scale \(c\). We evaluate the average diagnosis accuracy across all six scenarios described in Section \ref{sec:simulation} while varying one parameter at a time and maintaining the others at their default values (\(\tau_{\text{NCI}}=0.5\) dB, \(\alpha_{\text{sig}}=0.05\), \(c=2\)).

Fig. \ref{fig:sensitivity}a shows that the accuracy remains consistently high (above 85\%) for \(\tau_{\text{NCI}}\) ranging from 0.25 to 2.0 dB, indicating stable NMM detection performance. Fig. \ref{fig:sensitivity}b demonstrates that stricter significance levels (e.g., \(\alpha_{\text{sig}}=0.001\)) yield improved accuracy, suggesting that minimizing false positive SMM detections is beneficial. Fig. \ref{fig:sensitivity}c reveals that the performance is stable for small probing scales ($c \le 2.5$) but gradually decreases for larger values, confirming that moderate scaling ($c \approx 2$) is a good choice for distinguishing between optimism and pessimism. 

\begin{figure*}[!htb]
  \centering
  \includegraphics[width=\textwidth]{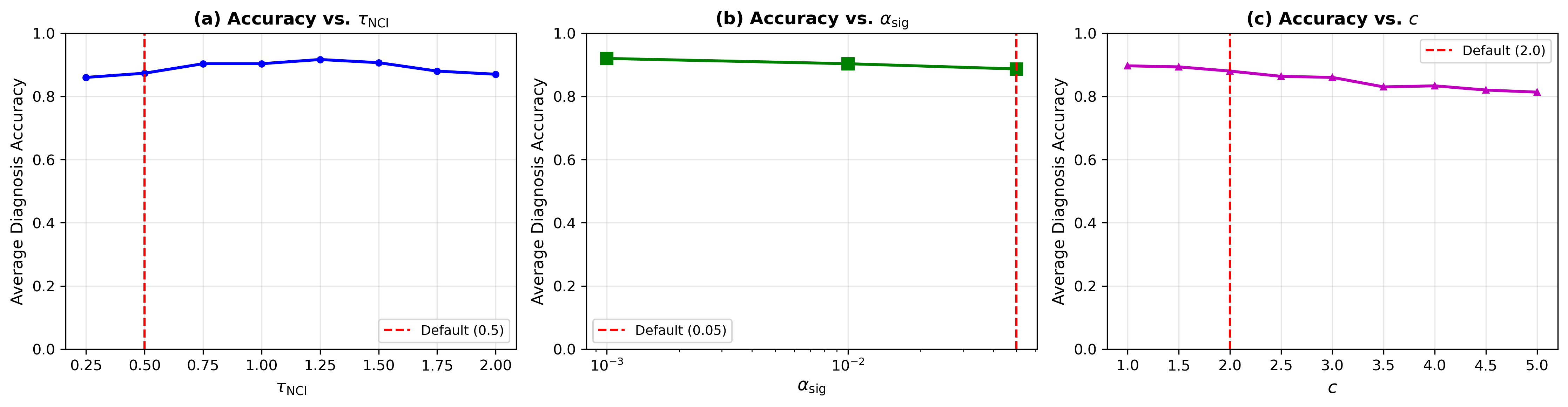}
  \caption{Sensitivity analysis of the diagnosis accuracy with respect to (a) NCI threshold \(\tau_{\text{NCI}}\), (b) significance level \(\alpha_{\text{sig}}\) for ELT, and (c) directional probe scaling \(c\). The vertical dashed lines indicate the default parameter values used in this work.}
  \label{fig:sensitivity}
\end{figure*}

\end{revision}

\begin{revision}
\subsection{Scalability Analysis}\label{sec:scalability}
To evaluate the applicability of the proposed framework to high-dimensional state estimation problems, we conducted additional Monte Carlo simulations with state dimensions $d$ ranging from 2 to 100. The experimental setup follows the same protocol as in Section \ref{a.-experimental-design}.

Fig. \ref{fig:scalability} illustrates the diagnosis accuracy across different state dimensions. The diagnosis accuracy remains consistently high (generally above 80\%) across all tested dimensions, with no significant degradation observed even at $d=100$. Notably, the detection of Pessimism+SMM exhibits exceptional stability, achieving near-perfect accuracy in high-dimensional settings. This consistent performance confirms the scalability of the proposed multi-metric framework for complex, high-dimensional navigation and tracking applications.

\begin{figure}[!htb]
  \centering
  \includegraphics[width=\linewidth]{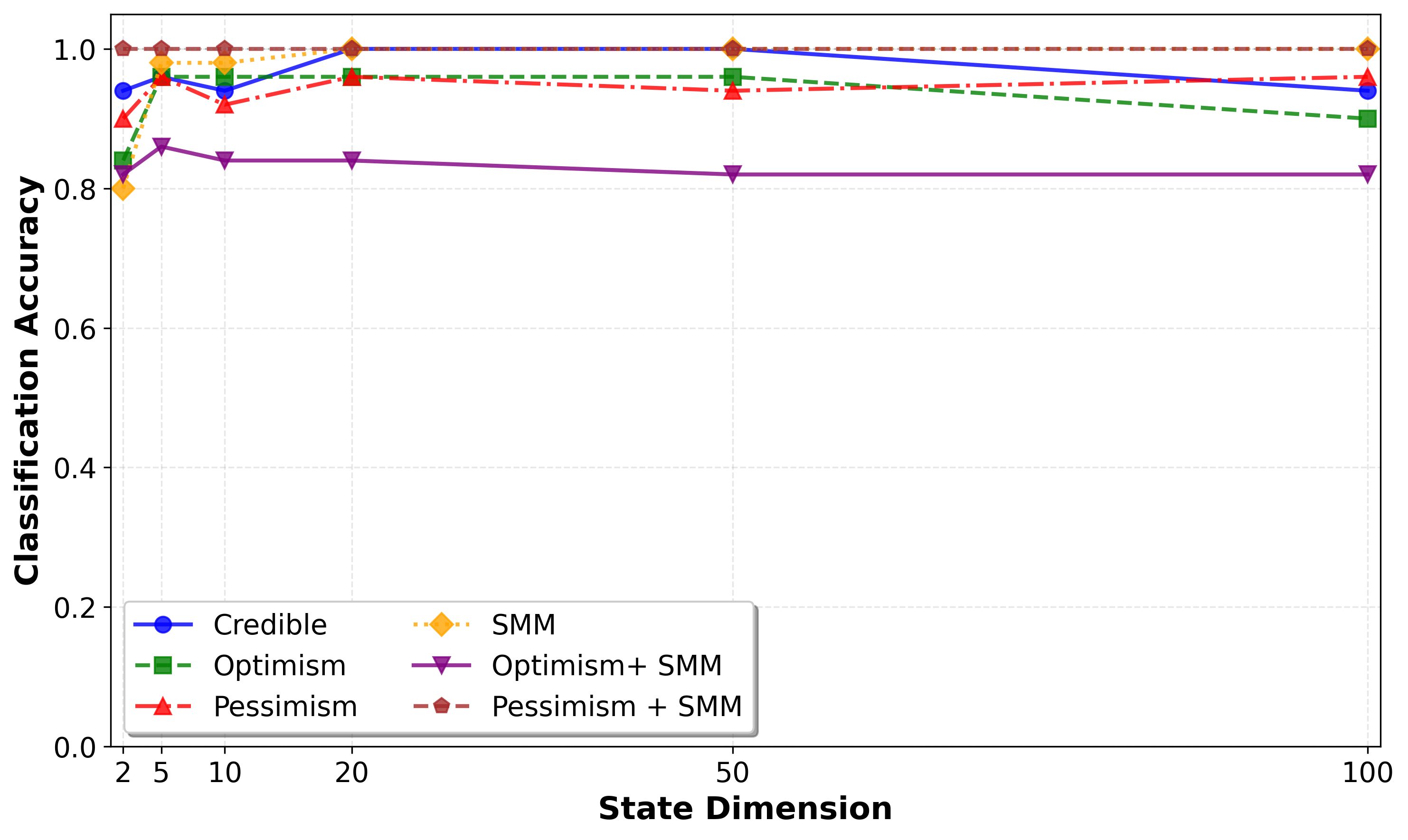}
  \caption{Diagnosis accuracy of the proposed method across varying state dimensions ($d=2$ to $d=100$) for six credibility scenarios.}
  \label{fig:scalability}
\end{figure}
\end{revision}

\section{UWB Positioning Experiment}\label{sec:uwb}
This section evaluates the proposed framework on the STAR-loc dataset \cite{dumbgen2023star}, a real-world dataset for stereo and range-based localization. Specifically, we use the UWB measurements collected under configuration s3 with landmark set v2 and grid trajectory (starloc\_data\_grid\_s3\_uwb.csv), which comprises range measurements between a mobile UWB tag and eight fixed anchors with known coordinates. The surveyed range from anchor to tag is also provided in the dataset, which enables the calculation of authentic UWB range measurement errors. \begin{revision}The ground-truth position of the UWB tag is provided by the Vicon motion capture system, as documented in the STAR-loc dataset. \end{revision}

\subsection{Data preprocessing in UWB Positioning Experiments}\label{app:preprocessing}
\begin{figure}[!htb]
  \centering
\subfloat[~~~~]{%
     \includegraphics[width=75mm]{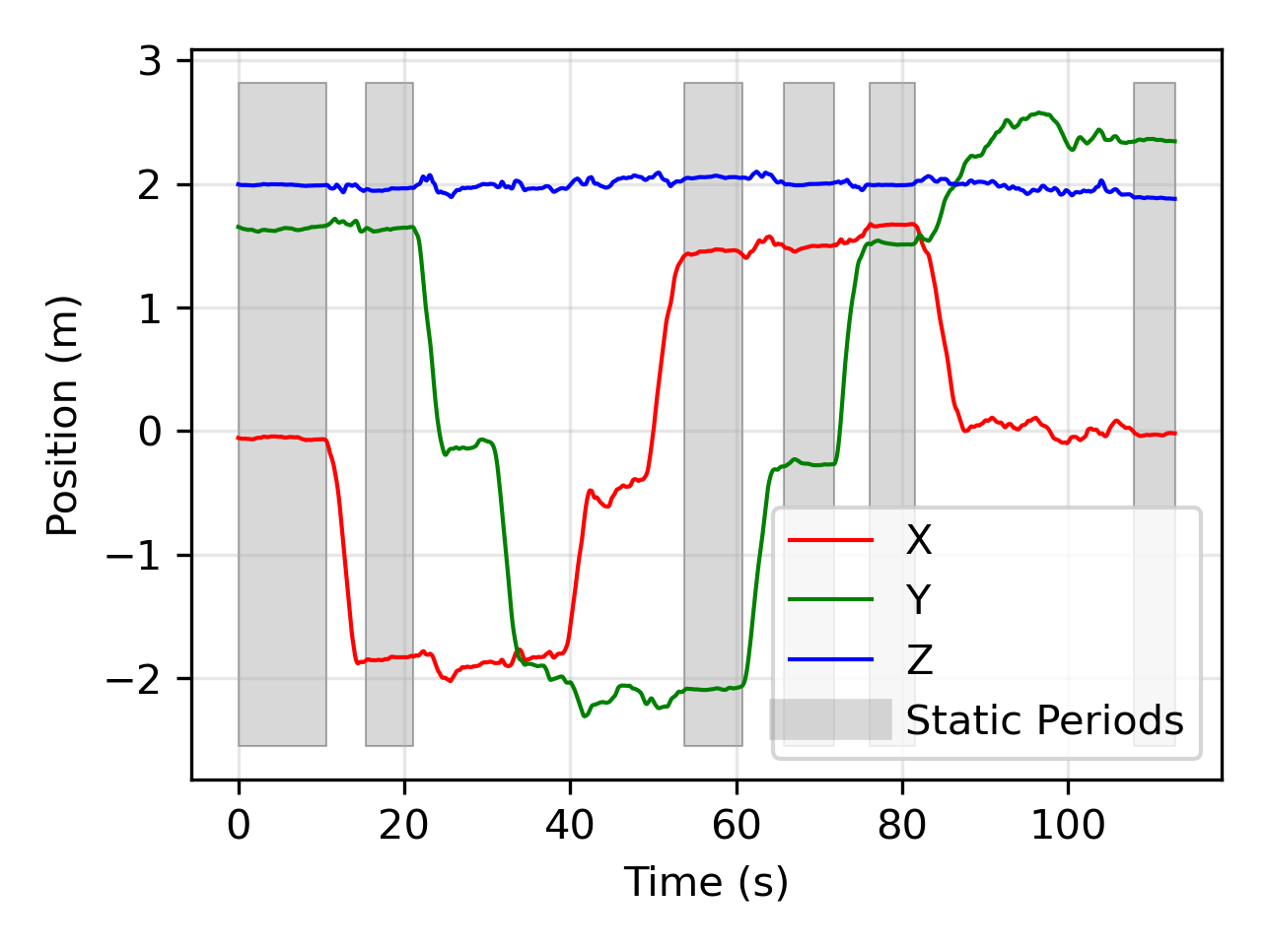}}
  \\
\subfloat[~~~~]{%
      \includegraphics[width=80mm]{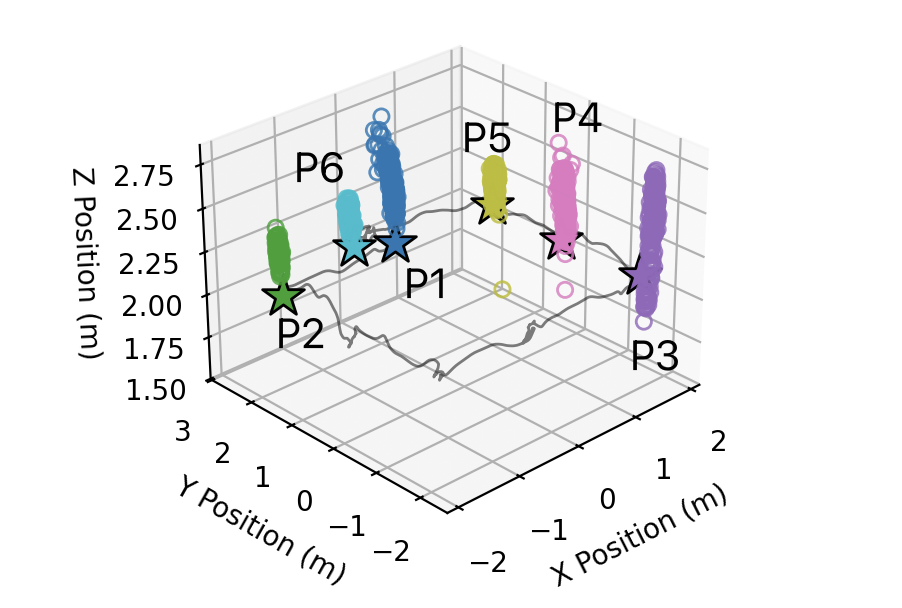}}
\caption{(a) The coordinates of the UWB tag against time, where the static period is marked as the shaded area; (b) \begin{revision}The 3-D trajectory of the UWB tag and the positioning solutions at six static locations. The `star' stands for ground-truth location, and the `circle' represents the positioning estimation.\end{revision}}
\label{fig:segmentation}
\end{figure}

\begin{figure*}[!htb]
  \centering
\subfloat[~~~~]{%
     \includegraphics[width=\textwidth]{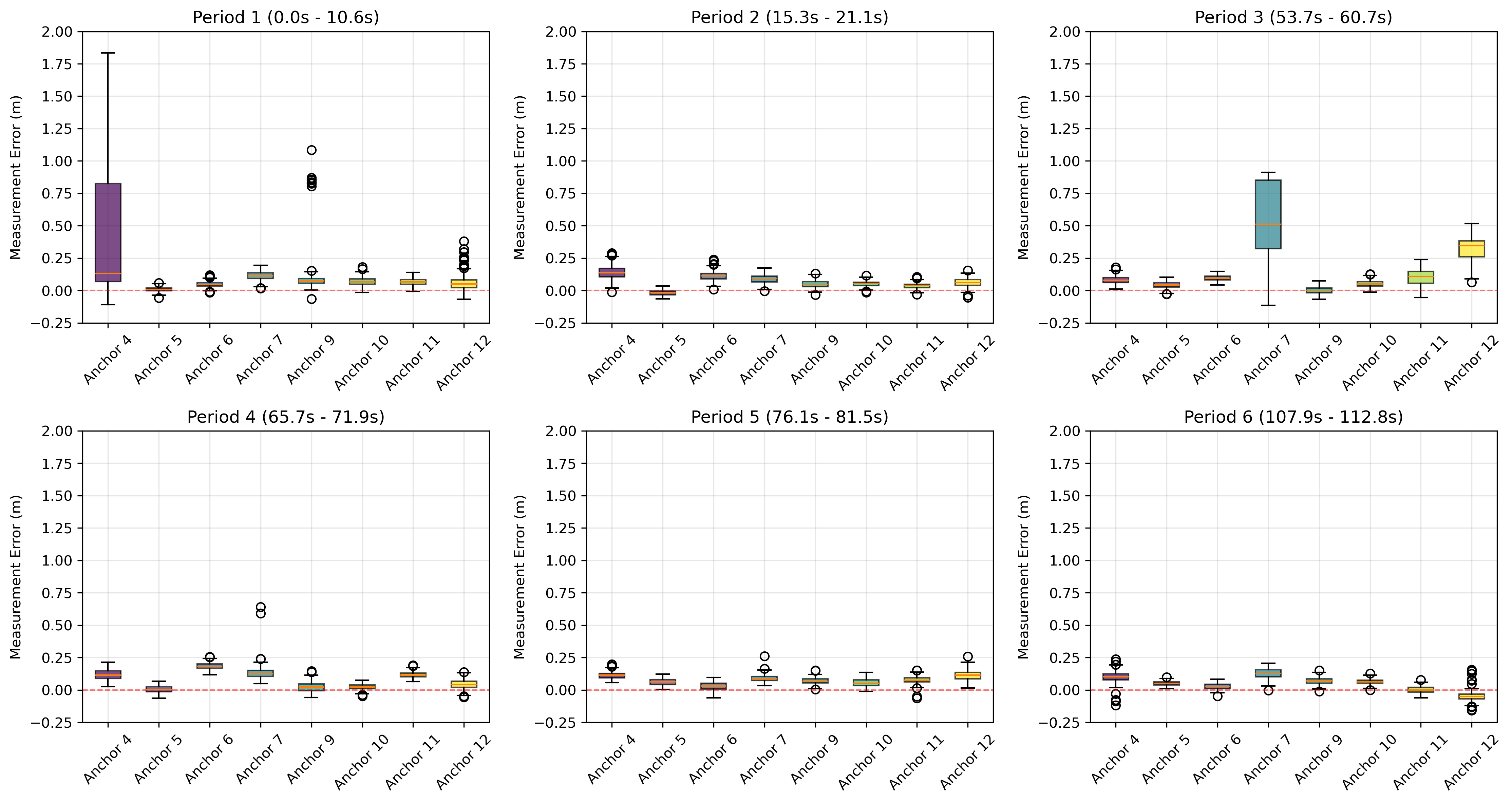}}
\caption{The distribution of UWB range measurement errors (related to each anchor) at each static period. }
\label{fig:error_distribution}
\end{figure*}

\begin{figure*}[!htb]
  \centering
\subfloat[~~~~]{%
      \includegraphics[width=\textwidth]{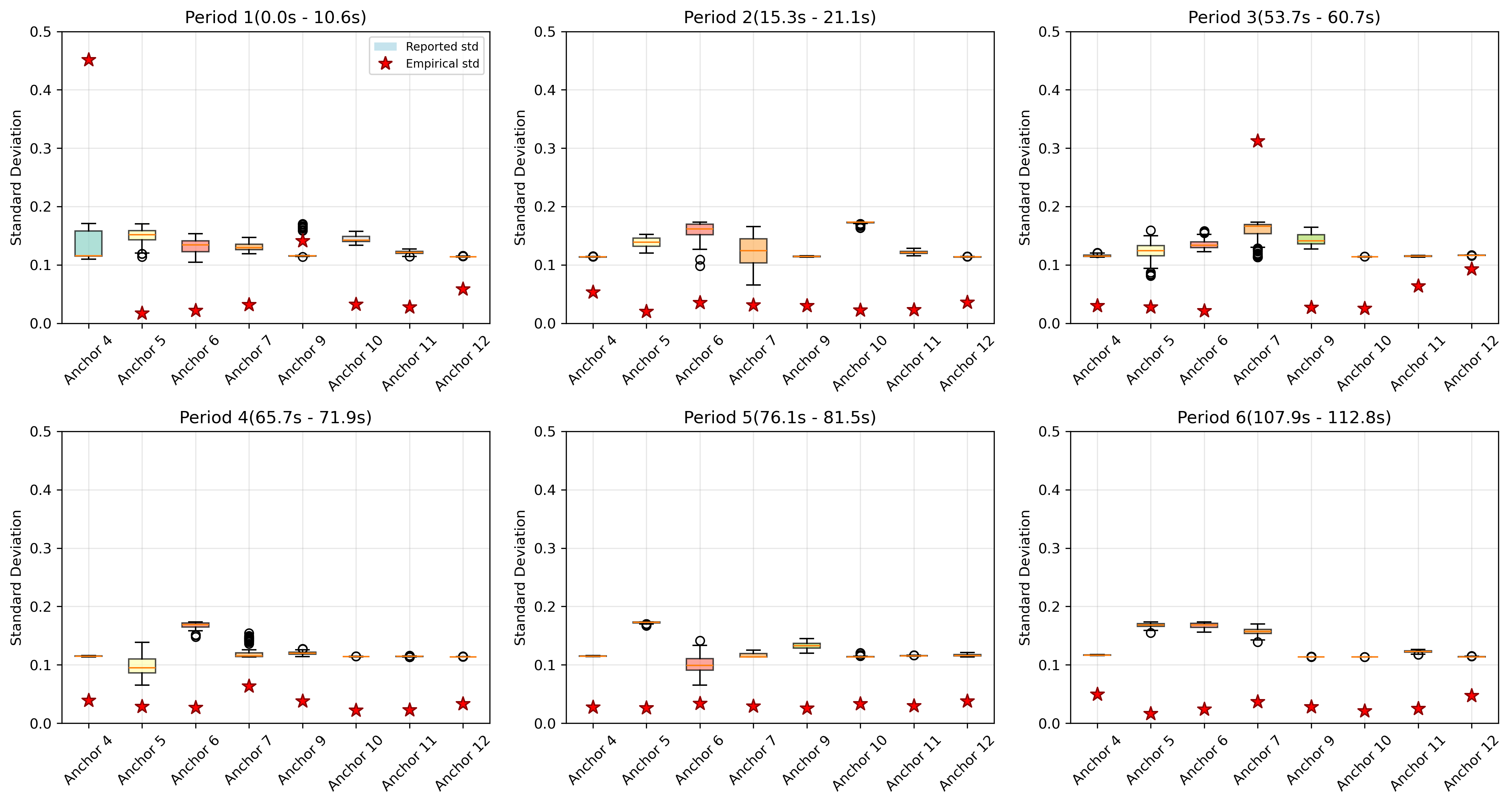}}
\caption{The distribution of reported standard deviations at each static period. The red stars indicate the empirical standard deviation, which is calculated using the UWB range measurement errors. }
\label{fig:std_distribution}
\end{figure*}

The data processing pipeline encompasses two primary stages. Initially, we implement a velocity-based segmentation strategy to identify static periods. We select periods where the tag's three-dimensional velocity remains below 0.1 m/s for a minimum duration of 4 seconds as static periods, ensuring stable measurement conditions for subsequent analysis. Fig. \ref{fig:segmentation} shows the segmentation of the dataset in terms of the xyz-coordinate series and 3D-trajectory. Fig. \ref{fig:error_distribution} plots the distribution of UWB range measurement errors with respect to each anchor, revealing significant systematic biases in measurements across all static periods. Fig. \ref{fig:std_distribution} shows boxplots of reported uncertainties (standard deviations) of UWB range measurements with respect to each anchor. The red stars indicate the empirical standard deviation, which is calculated using the UWB range measurement errors. The comparison clearly demonstrates that most reported measurement uncertainties are pessimistic.

\begin{revision}In the second stage, we apply a temporal aggregation strategy in each static period to address the challenge of asynchronous UWB measurements. Specifically, UWB range measurements collected within a 0.03-second window from distinct anchors are grouped into packets. For each packet, we calculate the 3-D position estimate. Specifically, the state vector is defined as $x_k = [p_{x,k}, p_{y,k}, p_{z,k}]^\top$, where $p_{x,k}$, $p_{y,k}$, and $p_{z,k}$ are the x-, y-, and z-coordinates of the UWB tag at period $k$. Since the two-way ranging (TWR) protocol is used when collecting the data \cite{dumbgen2023star}, the observation model for the $i$-th anchor is given by:
\begin{equation}
z_k^{(i)} = \|x_k - a^{(i)}\|_2 + v_k^{(i)},
\end{equation}
where $z_k^{(i)}$ denotes the range measurement from the $i$-th anchor at period $k$, $a^{(i)}$ is the coordinate of the $i$-th anchor, $v_k^{(i)}\sim\mathcal{N}(0, (\sigma_k^{(i)})^2)$ 
represents the measurement noise, and $\sigma_k^{(i)}, i=1,2,\cdots,m$ are the reported standard deviations in the dataset. The position estimation is obtained by applying the weighted least squares (WLS) method (in an iterative approach).
\end{revision}

This per-static-period approach ensures each positioning solution is computed using a consistent set of near-simultaneous measurements. Fig. \ref{fig:segmentation}b shows the positioning results in each static period. As can be seen, the positioning results do not surround the ground truth, indicating the presence of SMM. This is consistent with the findings in Fig. \ref{fig:error_distribution}. Given the observations in Figs. \ref{fig:error_distribution}, \ref{fig:std_distribution}, and \ref{fig:segmentation}b, an ideal credibility assessment method should identify the positioning estimation of such a dataset as both SMM and pessimism.

\subsection{Diagnosis Result Analysis}
\begin{revision}In implementing the credibility diagnosis, we use $\alpha_{\text{sig}} = 0.05$, $c = 2$, and $\tau_{\text{NCI}} = 0.5$ dB\end{revision}. 
Table \ref{tab:uwb_results} compares the credibility diagnosis results
between our proposed method and the two baseline approaches.
As predicted by our initial data analysis, the proposed method correctly
identifies the combination of pessimism and SMM in
periods 2-6. \begin{revision}As for period 1, it is diagnosed as SMM rather than ``Pessimism + SMM". One possible reason is that the reported standard deviation of Anchor 4's measurements is significantly smaller than the empirical standard deviation (optimism), contrasting with the general pessimism of other anchors' measurements. This strong optimism from Anchor 4's measurements effectively counteracts the pessimism contributed by the other anchors' measurements. Consequently, the covariance of the final position estimate appears balanced with respect to the noise model, masking the underlying NMM and leaving only the SMM to be detected. To verify this, we excluded Anchor 4 and repeated the diagnosis for Period 1, which yielded the diagnosis result ``Pessimism + SMM". This finding confirms that the proposed framework is designed to evaluate the credibility of the estimation result (the net effect of all inputs) rather than the credibility of individual measurements. \end{revision}

When looking at the results from the baseline methods, we find that these baseline methods failed to achieve the same complete diagnosis as the proposed method. Moreover, both the NCI and NEES based methods fail to identify the SMM in estimation. These results demonstrate that the
proposed method provides a more nuanced and complete credibility assessment compared to baseline methods, successfully capturing both the SMM and NMM revealed in the dataset.

\begin{table}[!htb]
  \caption{Diagnosis results for UWB positioning on STAR-loc dataset}
  \label{tab:uwb_results}
  \centering
  \begin{tblr}{
    hline{1,8} = {-}{1.5pt},
    hline{2} = {-}{0.75pt},
    colspec = {Q[5]Q[10]Q[10]Q[10]Q[10]},
  }
  Period ID & Labeled & Proposed           & NEES        &NCI  \\
  1 & Pessimism + SMM & SMM             & Optimism & Optimism   \\
  2 & Pessimism + SMM & Pessimism + SMM & Credible & Pessimism   \\
  3 & Pessimism + SMM & Pessimism + SMM & Optimism & Optimism \\
  4 & Pessimism + SMM & Pessimism + SMM & Credible & Pessimism \\
  5 & Pessimism + SMM & Pessimism + SMM & Credible & Pessimism \\
  6 & Pessimism + SMM & Pessimism + SMM & Credible  & Pessimism 
  \end{tblr}
\end{table}

\section{Discussion of computational complexity}\label{sec:computation}

\begin{revision}
In this section, we discuss the computational complexity of the proposed unified framework. The framework is composed of four key components: ELT, NCI, NLL, and ES. We analyze the computational complexity of each component and its impact on the diagnosis performance. All computations are performed on a laptop (MacBook Air with 16GB RAM and Apple M4 processor). 

\textbf{ELT:} The computational complexity of the ELT is dominated by the pairwise distance evaluations required by the energy-distance statistic in each sign-flip randomization, as shown in Eq. \eqref{eq:energy_distance}. For a batch of $N$ whitened errors $\{s_k\}_{k=1}^{N}$ and $B$ Monte-Carlo sign-flip iterations, each iteration evaluates a $U$-statistic $\left(\|s_i + s_j\|^\alpha - \|s_i - s_j\|^\alpha\right)$ over all $N(N-1)/2$ sample pairs, leading to an $O(N^2)$ cost per iteration and an overall complexity of $O(B\cdot N^2)$. As the computational cost of the ELT test scales linearly with $B$, we need to carefully consider the trade-off between $B$ and testing performance. 

\begin{figure*}[!htb]
  \centering
  \subfloat[~~~~]{%
     \includegraphics[width=60mm]{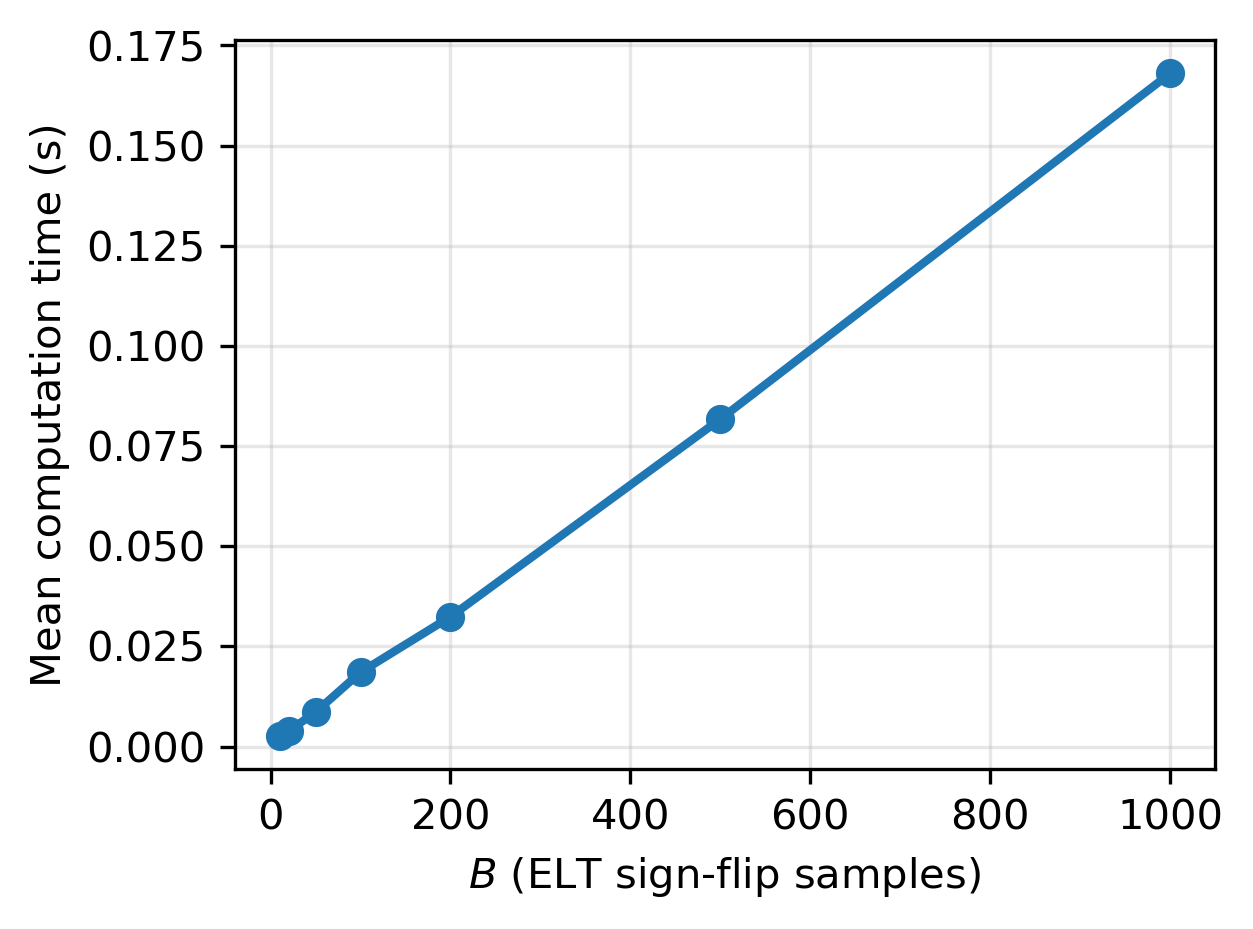}}
  \subfloat[~~~~]{%
     \includegraphics[width=60mm]{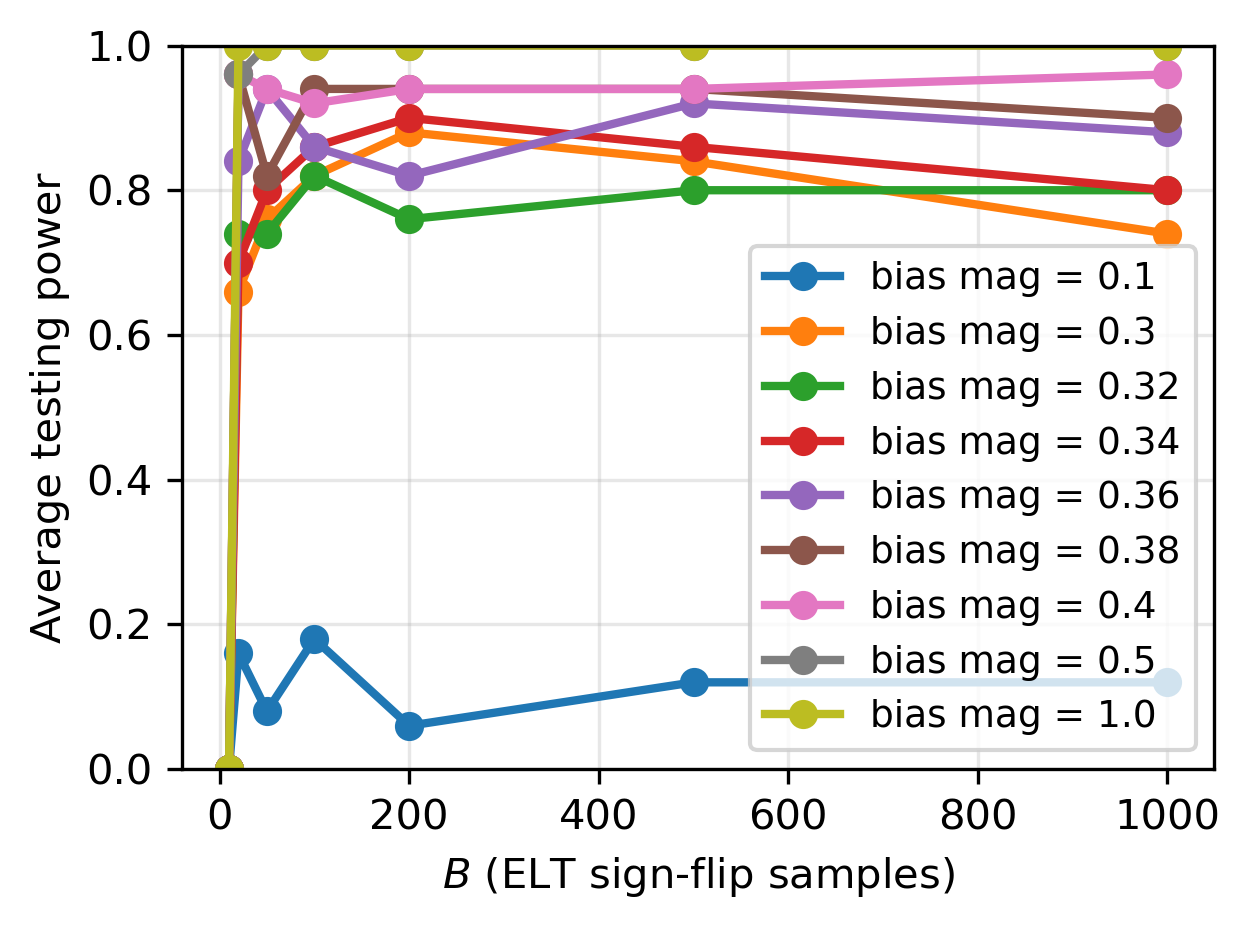}}
  \subfloat[~~~~]{%
    \includegraphics[width=60mm]{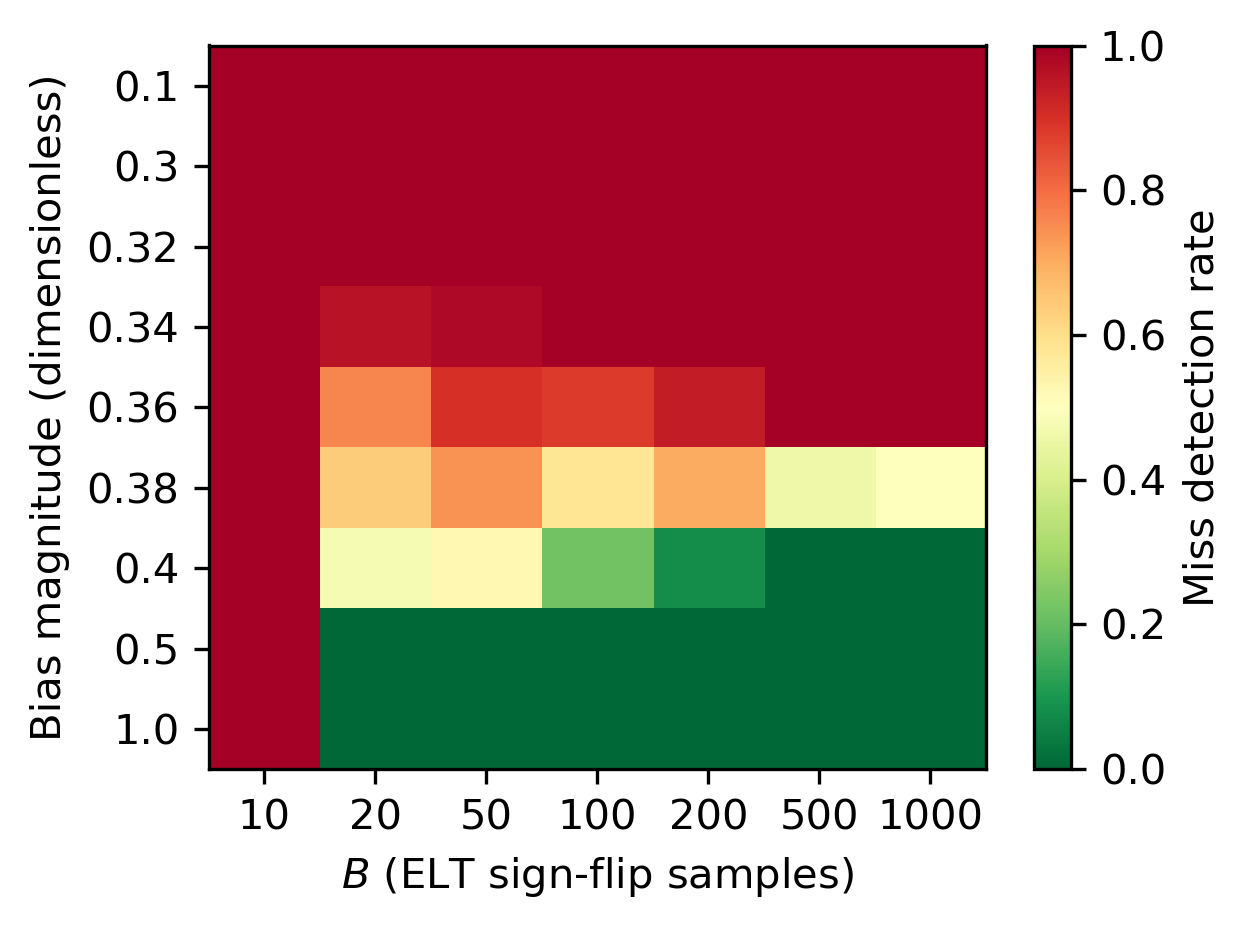}}
  \caption{(a) Mean computation time per ELT test versus the number of sign-flip samples $B$; (b) Average testing power ($1-\text{missed detection rate}$) versus $B$; (c) Reliability analysis of the ELT test by showing the heatmap of miss-detection rate as a function of $B$ and injected bias magnitude.}
  \label{fig:ELT_sensitivity}
\end{figure*}

To investigate this trade-off, we examine the average testing power of the ELT test in an SMM-only synthetic scenario. We consider a 2-D state estimation problem ($d=2$) where true states $x_k$ are generated from a Gaussian distribution with mean $\mathbf{0}$ and covariance $\Sigma_{\text{true}} = \mathbf{I}_2$. State estimates are generated as $\hat{x}_k \sim \mathcal{N}(x_k + \mu, \Sigma_{\text{true}})$, where the systematic error $\mu$ has a fixed magnitude $\|\mu\|$ (varied in the experiment) and a random direction. The claimed covariances are calibrated, i.e., $\hat{\Sigma}_k = \Sigma_{\text{true}}$. We perform 50 independent trials for each pair of $(B, \|\mu\|)$, with each trial comprising $N=100$ samples. The significance level is set to $\alpha_{\text{sig}}=0.05$. 

Fig. \ref{fig:ELT_sensitivity}a reports the mean computation time per ELT test against $B$. This result confirms an approximately linear growth in runtime with respect to $B$, as predicted by the $O(B\cdot N^2)$ complexity. Fig. \ref{fig:ELT_sensitivity}b shows that the average testing power ($1-\text{missed detection rate}$) saturates quickly, where the gain from increasing $B$ becomes marginal once $B\ge 100$. This empirical observation aligns with the theory of permutation tests. According to Dwass (1957) \cite{dwass1957}, the "power loss" of a permutation test with $B$ iterations relative to an infinite number of permutations is approximately: $\text{Power Loss} \approx \frac{1}{B}$. Increasing $B$ from 100 to 1000 only reduces the theoretical power loss from $1\%$ to a mere $0.1\%$.

However, average testing power does not capture the reliability of the ELT decision. To evaluate this, we apply the ELT test to a fixed synthetic dataset. The experiment setting is almost the same as the testing-power experiment, except that for each $\|\mu\|$, we generate the synthetic dataset once and then apply the ELT test to it 50 times with different random seeds. Fig. \ref{fig:ELT_sensitivity}c plots the heatmap of the miss-detection rate as a function of $(B,\|\mu\|)$. When $\|\mu\|$ is small, the missed detection rate is very large at low $B$, indicating the test is unstable. As $B$ increases toward 1000, the decision becomes more stable. This demonstrates that a higher $B$ is primarily required for reliability rather than increasing the testing power.

Based on the above, we recommend $B\geq 1000$ for offline use cases (e.g., estimator development and supervised parameter tuning), where computational constraints are less stringent and high reliability is preferred, especially for subtle SMM. For online/real-time applications, a smaller $B$ (e.g., $B=100$) is typically sufficient when the SMM is pronounced (e.g., strong NLOS-induced biases in GNSS positioning). If high-precision online detection of small biases is required, one may increase $B$ and adopt acceleration strategies such as parallelization on multi-core CPU/GPU.

\textbf{ES:} In the calculation of ES, we approximate the expectation operation in Eq. \eqref{eq:es} using Monte Carlo sampling with $M$ samples. The computational complexity is dominated by the sample generation process, which involves Cholesky decomposition ($O(d^3)$) and matrix-vector multiplications ($O(M \cdot d^2)$). The remaining operations, including distance calculations and mean computation, scale linearly with $M$ and $d$. Thus, the overall complexity is $O(M \cdot d^2)$ when $M \gg d$, which is usually the case in practice. \par
To investigate the trade-off between computational cost and approximation accuracy, we evaluated the ES computation time and the approximation error across a range of $M$ values. The approximation error is quantified as the mean absolute difference between the ES calculated with $M$ samples and a high-precision reference value calculated with $M=5000$ samples. As shown in Fig. \ref{fig:ES_computation}a, the computation time of ES exhibits a linear increase with the number of samples $M$. Fig. \ref{fig:ES_computation}b demonstrates the convergence of the ES approximation. The difference relative to the reference value decreases significantly as $M$ increases. Notably, when $M \ge 500$, the variation in the difference becomes minimal (stabilizing around 0.04), suggesting that further increasing $M$ yields diminishing returns in accuracy. \par
Based on these findings, we conclude that setting $M$ between 500 and 1000 is an optimal choice for most applications, offering a favorable balance between accuracy and computational load. For scenarios demanding higher precision where computational resources permit, a range of $M=1000$--2000 may be adopted.

\begin{figure}[!htb]
  \centering
  \subfloat[~~~~]{%
     \includegraphics[width=75mm]{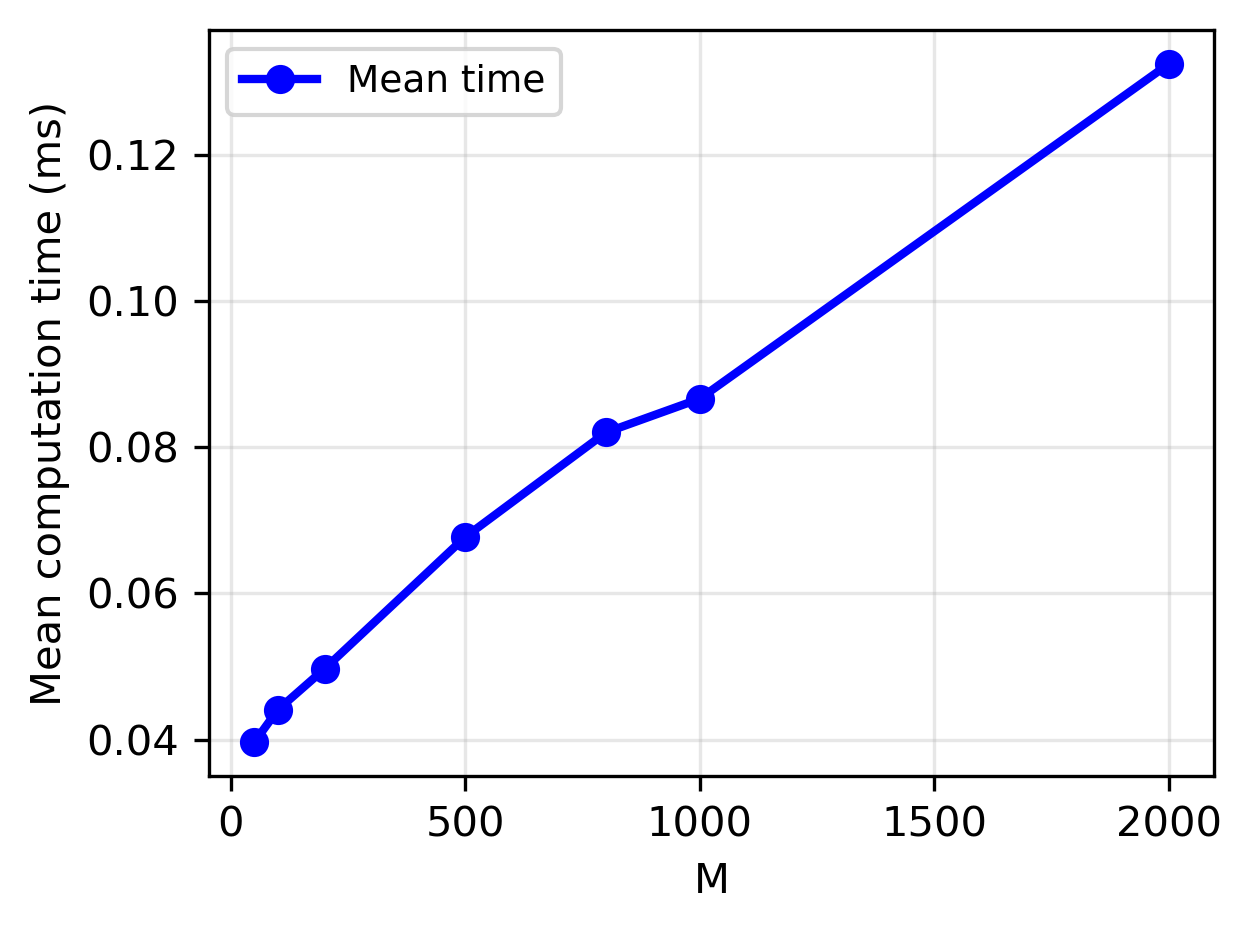}}
     \\
  \subfloat[~~~~]{%
     \includegraphics[width=75mm]{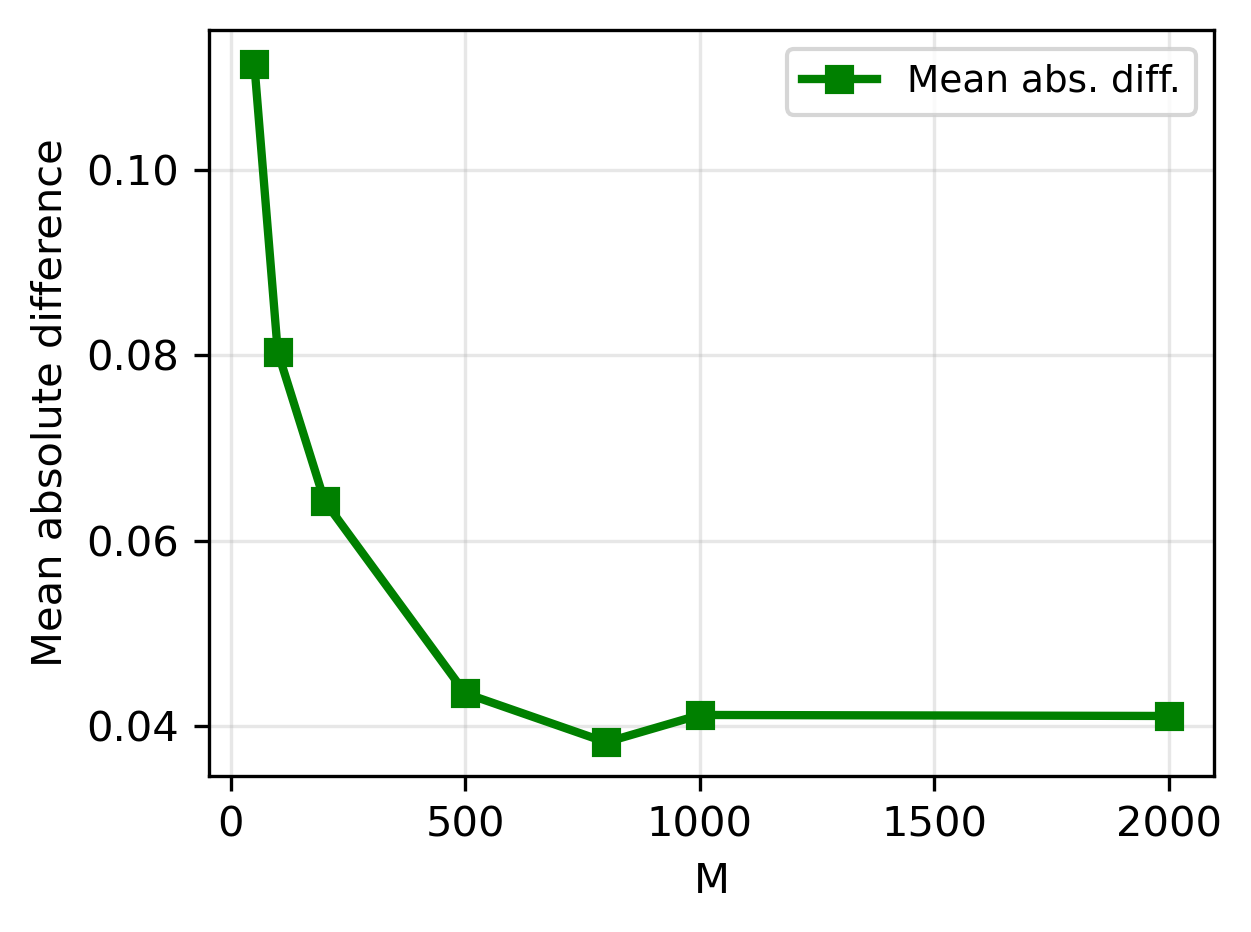}}
  \caption{(a) Mean computation time of ES versus the number of samples used in Monte Carlo sampling $M$; (b) Approximation error (mean absolute difference to the reference value) of ES versus $M$.}
  \label{fig:ES_computation}
\end{figure}

\textbf{NLL:} As shown in Eq. \eqref{eq:nll}, the computation of NLL is simply evaluating the log-likelihood function of the predictive distribution $\hat{F}_k$. When $\hat{F}_k\sim\mathcal N(\hat{x}_k,\hat{\Sigma}_k)$, the NLL is given by:
\begin{equation}
\frac{1}{2}\log((2\pi)^d |\hat{\Sigma}_k|) + \frac{1}{2}(x_k - \hat{x}_k)^\top \hat{\Sigma}_k^{-1} (x_k - \hat{x}_k).
\end{equation}
The computation of NLL involves the determinant and inverse of the covariance matrix $\hat{\Sigma}_k$, which typically requires Cholesky decomposition or similar matrix factorizations. For a $d$-dimensional state, these operations have a computational complexity of $O(d^3)$. 

\textbf{NCI:} The NCI calculation involves the computation of the NEES for each sample, which requires the inverse of the covariance matrix, as shown in Eq. \eqref{eq:nees}. For a $d$-dimensional state, the matrix inversion has a complexity of $O(d^3)$. 

\textbf{Overall Complexity:} Based on the execution flow of Algorithm \ref{alg:1}, the overall computational complexity is determined by the specific diagnosis path taken. The algorithm invariably executes the ELT test and NCI calculation. The ES and NLL probes are only triggered in the worst-case scenario (i.e., when SMM is detected and NCI suggests potential optimism). Therefore, the overall worst-case computational complexity can be expressed as:
\begin{equation}
\mathcal{C}_{\text{total}} = \underbrace{O(B \cdot N^2)}_{\text{ELT}} + \underbrace{O(N \cdot d^3)}_{\text{NCI \& NLL}} + \underbrace{O(N \cdot M \cdot d^2)}_{\text{ES}}.
\end{equation}
For offline diagnosis where $N$ is large, the quadratic term $O(B \cdot N^2)$ from ELT tends to dominate. For online or sliding-window applications (small $N$), the complexity is primarily driven by the Monte Carlo sampling in ES ($O(N \cdot M \cdot d^2)$).
\end{revision}

\section{Conclusions}\label{sec:conclusion}

In this work, we introduced a unified framework for
evaluating the credibility of state estimators by comprehensively integrating a suite of metrics, including NCI, NLL, and ES. Experimental results from  Monte Carlo simulations and real-world UWB positioning experiments 
confirmed the superiority of our multi-metric approach. The proposed method achieved high
classification accuracy across a range of challenging scenarios, whereas traditional single-metric methods show unreliable results. This work offers
practitioners a powerful tool to not only validate estimator performance
but also to diagnose specific modeling failures.

\appendices

\section{Non-negativity of NCI under SMM}\label{appendix-nci}

\textbf{Step 1: Simplify the Matrix Inverse}

Let $\hat{\Sigma}_k= \Sigma_k$  and $\mu_k \neq 0$, we have
$$
NCI(\{\hat{x}_k\}) = \frac{10}{N}\sum_{k=1}^N\log_{10}\frac{e_k^T \Sigma_k^{-1} e_k}{e_k^T (\Sigma_k + \mu_k \mu_k^T)^{-1} e_k} \,.
$$

The denominator contains the inverse of a sum of a matrix and a rank-1
matrix, which can be simplified using the Woodbury matrix
identity:
\begin{equation}(\Sigma_k + \mu_k \mu_k^T)^{-1} = \Sigma_k^{-1} - \Sigma_k^{-1}\mu_k(1+\mu_k^T\Sigma_k^{-1}\mu_k)^{-1}\mu_k^T\Sigma_k^{-1}\,.\end{equation}Since
\((1+\mu_k^T\Sigma_k^{-1}\mu_k)\) is a scalar, we can write the inverse
as:\begin{equation}(\Sigma_k + \mu_k \mu_k^T)^{-1} = \Sigma_k^{-1} - \frac{\Sigma_k^{-1}\mu_k\mu_k^T\Sigma_k^{-1}}{1+\mu_k^T\Sigma_k^{-1}\mu_k}\,.\end{equation}

\textbf{Step 2: Simplify the
Denominator}

Now, substitute this simplified inverse into the denominator of the
original expression:
\begin{equation}e_k^T (\Sigma_k + \mu_k \mu_k^T)^{-1} e_k = e_k^T \left( \Sigma_k^{-1} - \frac{\Sigma_k^{-1}\mu_k\mu_k^T\Sigma_k^{-1}}{1+\mu_k^T\Sigma_k^{-1}\mu_k} \right) e_k\end{equation}
\begin{equation} = e_k^T\Sigma_k^{-1}e_k - \frac{e_k^T\Sigma_k^{-1}\mu_k\mu_k^T\Sigma_k^{-1}e_k}{1+\mu_k^T\Sigma_k^{-1}\mu_k}\,.\end{equation}
Since \(e_k^T\Sigma_k^{-1}\mu_k\) and \(\mu_k^T\Sigma_k^{-1}e_k\) are
scalars and are transposes of each other, they are equal. The denominator can be simplified to:
\begin{equation}\frac{(e_k^T\Sigma_k^{-1}e_k)(1+\mu_k^T\Sigma_k^{-1}\mu_k) - (e_k^T\Sigma_k^{-1}\mu_k)^2}{(1+\mu_k^T\Sigma_k^{-1}\mu_k)}\,.\end{equation}

\textbf{Step 3: Simplify the Overall
Expression}

Substituting the simplified denominator into the NCI expression, the finalized NCI is given by:
\begin{equation}\log_{10}\left(\frac{(e_k^T\Sigma_k^{-1}e_k)(1+\mu_k^T\Sigma_k^{-1}\mu_k)}{(e_k^T\Sigma_k^{-1}e_k)(1+\mu_k^T\Sigma_k^{-1}\mu_k) - (e_k^T\Sigma_k^{-1}\mu_k)^2}\right)\,.\end{equation}

Given that \(\mu_k \neq 0\), it follows that
\(\mu_k^T\Sigma_k^{-1}\mu_k\) is always positive, and
\(e_k^T\Sigma_k^{-1}e_k\) is non-negative. We can use the Cauchy-Schwarz inequality for the inner product
defined by \(\Sigma_k^{-1}\):
\begin{equation}(e_k^T\Sigma_k^{-1}\mu_k)^2 \le (e_k^T\Sigma_k^{-1}e_k)(\mu_k^T\Sigma_k^{-1}\mu_k)\,.\end{equation}Therefore, the denominator is always
positive. It is obvious that the numerator is always greater
than or equal to the denominator, because $(e_k^T\Sigma_k^{-1}\mu_k)^2$ is
non-negative. Therefore, the term in the bracket is always equal to or larger than 1, and thus NCI is always non-negative when \(\mu_k \neq 0\). 

\section{Asymmetric Properties of NLL}\label{appendix-nll}
Eq. \eqref{eq:exp_nll_1} contains the term \(1/\rho + \ln(\rho) > 0\). For \(\rho > 1\), we have \(1/\rho + \ln(\rho) > 0\) and
\(\rho - \ln(\rho) > 0\). The difference between these two
expressions is:\begin{equation} \Delta(\rho) = (\rho - \ln(\rho)) - \left(\frac{1}{\rho} + \ln(\rho)\right) = \rho - \frac{1}{\rho} - 2\ln(\rho) \,.\end{equation}To
determine the sign of \(\Delta(\rho)\) for \(\rho > 1\), we calculate the derivative of \(\Delta(\rho)\) with respect to \(\rho\):\begin{equation} \frac{d\Delta}{d\rho} = 1 + \frac{1}{\rho^2} - \frac{2}{\rho} = \left(1 - \frac{1}{\rho}\right)^2 \,.\end{equation}
For \(\rho > 1\), the term \((1 - 1/\rho)\) is positive, which implies
that \((1 - 1/\rho)^2 > 0\). Thus, \(d\Delta/d\rho > 0\), and
\(\Delta(\rho)\) is a strictly increasing function for \(\rho > 1\).

At the boundary point \(\rho=1\), we have:
\begin{equation}\Delta(1) = 1 - \frac{1}{1} - 2\ln(1) = 0\,.\end{equation} Since
\(\Delta(\rho)\) is increasing for \(\rho > 1\) and \(\Delta(1)=0\), it
follows that \(\Delta(\rho) > 0\) for all \(\rho > 1\). This proves that
for any \(\rho > 1\): \begin{equation}\rho - \ln(\rho) > \frac{1}{\rho} + \ln(\rho)\,.\end{equation}
Therefore,
\begin{equation}|\mathbb{E}[\mathrm{NLL}](\rho)| < |\mathbb{E}[\mathrm{NLL}](1/\rho)|\,.\end{equation}

\section{Non-negativity of ES}\label{appendix-es}

The energy distance is a metric for
  measuring the distance between two probability distributions. For two
  independent random variables \(X\) and \(Y\), the energy distance is
  defined as:
  \begin{equation}\mathcal{E}(X,Y) = 2\mathbb{E}\|X-Y\|_2 - \mathbb{E}\|X-X'\|_2 - \mathbb{E}\|Y-Y'\|_2\end{equation}
  where \(X, X'\) are independent and identically distributed (i.i.d.)
  and \(Y, Y'\) are i.i.d. A key property of the energy distance is that it is always non-negative. \(\mathcal{E}(X,Y) \ge 0\), and
  \(\mathcal{E}(X,Y) = 0\) if and only if the distributions of \(X\) and
  \(Y\) are identical.

  The energy score is a
  special case of the energy distance. We can interpret the true
  observed outcome \(x_k\) as a degenerate probability distribution---a
  Dirac measure, which is a distribution that places all its
  probability mass at a single point, \(x_k\). Let's call this
  distribution \(G_k\).

  Let the predictive distribution be
  \(F_k\) and the true distribution be the Dirac measure \(G_k\). We can
  now express the energy distance between \(F_k\) and \(G_k\):
  
  \begin{equation}
  \begin{aligned}
  \mathcal{E}(F_k, G_k) = &2\mathbb{E}_{\substack{Y\sim F_k \\ X\sim G_k}}\|Y-X\|_2 - \mathbb{E}_{\substack{Y,Y'\sim F_k \\ \text{i.i.d.}}}\|Y-Y'\|_2 \\
  &- \mathbb{E}_{\substack{X,X'\sim G_k \\ \text{i.i.d.}}}\|X-X'\|_2 \,.
  \end{aligned}
  \end{equation} 
  The first term is \(\mathbb{E}\|Y-X\|_2\). Since \(X\) is always
    equal to \(x_k\), this simplifies to
    \(\mathbb{E}_{Y\sim F_k}\|Y-x_k\|_2\). The second term, \(\mathbb{E}\|Y-Y'\|_2\), is identical to the
    second term of the ES definition. The third term is \(\mathbb{E}\|X-X'\|_2\). Since both \(X\) and
    \(X'\) are always equal to \(x_k\), their distance is always zero,
    so this term is \(0\). Therefore, the energy distance can be simplified as:
  \begin{equation}\mathcal{E}(F_k, G_k) = 2\mathbb{E}_{Y\sim F_k}\|Y-x_k\|_2 - \mathbb{E}_{Y,Y'\sim F_k}\|Y-Y'\|_2 \,.\end{equation}
  By comparing this to the ES definition, we can see that:
  \begin{equation}ES(F_k,x_k)=\tfrac{1}{2}\mathcal{E}(F_k, G_k) \,.\end{equation}

It is known that the energy distance \(\mathcal{E}(F_k, G_k)\) is always
non-negative; and therefore, the energy score, which is half of the
energy distance, must also be non-negative. The minimum value of the energy score is \(0\), and this occurs when the
predictive distribution \(F_k\) perfectly matches the true distribution
\(G_k\) (i.e., when the model predicts the exact true outcome with
\(100\%\) certainty).

\bibliographystyle{IEEEtran}
\bibliography{ref} 

\end{document}